\documentclass[jap,preprint,letterpaper,showpacs,amsmath,amssymb,aip,export]{revtex4-2}

\usepackage{braket}
\usepackage{graphicx}
\usepackage{dcolumn}
\usepackage{bm}
\newlength{\figwidth}
\begin{document}

\title{Sequential Bayesian experiment design for adaptive Ramsey sequence measurements.}

\author{Robert D. McMichael}
\affiliation{Physical Measurement Laboratory, National Institute of Standards and Technology, Gaithersburg, Maryland 20899, USA}
\email{rmcmichael@nist.gov}

\author{Sergey Dushenko}
\affiliation{Physical Measurement Laboratory, National Institute of Standards and Technology, Gaithersburg, Maryland 20899, USA}
\affiliation{Institute for Research in Electronics and Applied Physics, University of Maryland, College Park, Maryland 20742, USA}

\author{Sean M. Blakley}
\affiliation{Physical Measurement Laboratory, National Institute of Standards and Technology, Gaithersburg, Maryland 20899, USA}

\date{\today}
\homepage{https://www.overleaf.com/project/5fb3d77e76b5462da1f9bed6}

\begin{abstract}
The Ramsey sequence is a canonical example of a quantum phase measurement for a spin qubit. In Ramsey measurements, the measurement efficiency can be optimized through careful selection of settings for the phase accumulation time setting, $\tau$. This paper implements a sequential Bayesian experiment design protocol for the phase accumulation time in low-fidelity Ramsey measurements, and performance is compared to both a previously reported adaptive heuristic protocol and random setting choices. A workflow allowing measurements and design calculations to run concurrently largely eliminates computation time from measurement overhead. When precession frequency is the lone parameter to estimate, the Bayesian design is faster by factors of 2 and 4 relative to the adaptive heuristic and random protocols respectively. 
\end{abstract}
\maketitle

\section{Introduction}

The development of diamond NV centers as measurement tools is one of the most significant achievements of applied quantum physics in recent years. From initial concept\cite{chernobrod_spin_2005} and first demonstrations, \cite{balasubramanian_nanoscale_2008, maze_nanoscale_2008, taylor_high-sensitivity_2008} NV-based magnetometry measurements\cite{rondin_magnetometry_2014} 
have been developed for condensed-matter physics,
\cite{hingant_measuring_2015, van_der_sar_nanometre-scale_2014, tetienne_nanoscale_2014, tetienne_quantum_2017, rondin_stray-field_2013} 
engineering,\cite{turner_magnetic_2020, horsley_microwave_2018, appel_nanoscale_2015} 
biology\cite{kucsko_nanometre-scale_2013, mcguinness_quantum_2011, mccoey_quantum_2020, ziem_highly_2013, ermakova_detection_2013, barry_optical_2016}, 
nanoscale nuclear magnetic resonance, \cite{devience_nanoscale_2015, rugar_proton_2014, haberle_nanoscale_2015, mamin_nanoscale_2013, bhallamudi_nitrogen-vacancy_2015, staudacher_nuclear_2013} 
and commercial instrumentation is now becoming available.

The qubit that draws attention to the NV center is the $S = 1$ spin of the electronic ground state, which forms around a nitrogen atom, neighboring vacancy, and trapped electron in a diamond crystal. The coherence time of the spin state can extend into the millisecond range.\cite{jahnke_long_2012, balasubramanian_ultralong_2009} At ambient temperatures, the spin state can be initialized and read out using laser light and detection of emitted photons. With incident green light, an NV center will begin to cycle between its electronic ground state and an excited state, absorbing a photon and relaxing.  The key factor is that the relaxation process in NV centers is spin-dependent.  Centers excited from the $m_s = 0$ state will relax by emitting a photon and return to the same $m_s = 0$ state, but centers excited from the $m_{\rm s} = \pm1$ states can also relax without emitting a photon, and may switch one-way from $m_{\rm s} = \pm 1$ to $m_s = 0$ in the process. Because readout resets the state to $m_{\rm s} = 0$, only photons from the first few absorb-relax cycles carry information about the initial spin state. 

A persistent problem in NV center measurements is that it is often difficult to efficiently collect these few meaningful, spin-dependent photons. Improvements in collection efficiency have been demonstrated using 
solid immersion lenses\cite{hadden_strongly_2010}, 
optical resonators,\cite{jensen_cavity-enhanced_2014, riedrich-moller_nanoimplantation_2015} 
and fabricated diamond nanostructures.\cite{babinec_diamond_2010, maletinsky_robust_2012, pelliccione_scanned_2016}
Other characteristics of NV centers have also been successfully exploited. Additional information can be gleaned from the timing of emitted photons\cite{dinani_bayesian_2019}.  More exotic approaches include spin-dependent ionization of the NV center into different long-lasting charge states where differences in emission spectra can be measured through many excitation-emission cycles.\cite{waldherr_violation_2011, aslam_photo-induced_2013, shields_efficient_2015, bluvstein_identifying_2019, jaskula_improved_2019, hacquebard_charge-state_2018, hopper_near-infrared-assisted_2016} Another approach has been to exploit features of a metastable state in the ``dark'' relaxation path of $m_s = \pm1$ states.\cite{acosta_optical_2010, acosta_broadband_2010}

While these methods demonstrate improvements to readout fidelity, additional gains can be achieved through efficient measurement design. More efficient measurement designs allow experiments to measure faster and/or more precisely. In quantum computing, Shor's algorithm includes an important example of efficient readout of spin phase at the Heisenberg limit where frequency uncertainty $\Delta_{\omega_0}$ scales as $\Delta_{\omega_0} \propto T^{-1}$ for measurement time $T$. For some qubits, adaptive Ramsey measurements have exhibited Heisenberg scaling.\cite{higgins_entanglement-free_2007,berry_how_2009,said_nanoscale_2011,cappellaro_spin-bath_2012,dinani_bayesian_2019}

In many cases, however, the readout is dominated by classical readout noise. 
The purpose of this paper is to compare measurement designs for Ramsey measurements in the low-fidelity, averaged readout regime, with emphasis on sequential Bayesian experiment design. Section \ref{sec:background} reviews Ramsey measurements and introduces sequential Bayesian experiment design.  Section \ref{sec:results} presents the results of measurement simulations comparing different protocols, showing substantial improvement of the Bayesian method over the other designs.

\section{Background\label{sec:background}}

\subsection{Ramsey experiment}
\begin{figure}[tb]
    \centering
    \includegraphics[width=\figwidth]{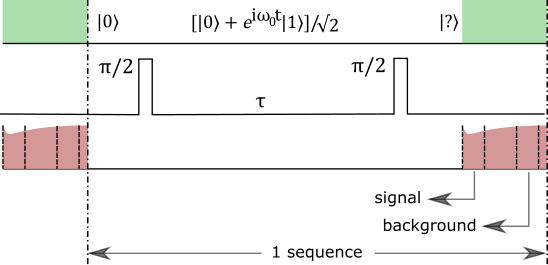}
    \caption{Timing diagram of Ramsey pulse sequences. Green areas in the top line denote green laser excitation. In the middle line, a microwave $\pi$/2 pulses initiates precession. After a time $\tau$ selected by the experiment design, another $\pi$/2 pulse projects the state for readout. Red areas in the bottom line show time intervals where the NV center emits red photons. Dashed lines show collection windows of the photons from the NV center with signal photons counted early in the laser pulse, and background photons counted during re-initialization. The horizontal axis corresponds to time passed during the experiment.}
    \label{fig:figure1}
\end{figure}

The Ramsey sequence is an example of quantum interferometry, primarily used to measure the energy difference $\hbar\omega_0$ between two states. When the energy difference depends on a signal, the Ramsey sequence is the basis of quantum sensing\cite{degen_quantum_2017, taylor_high-sensitivity_2008}. In section \ref{sec:results}, results are presented in terms of magnetometry, where the signal is magnetic field, and the energy difference is due to the Zeeman interaction. 

The Ramsey sequence is illustrated in figure \ref{fig:figure1}. First, the spin of the NV's electronic ground is initialized to $\ket{0}$ with a few microseconds of green light illumination followed by time to allow optically excited states to relax.  Next, the spins are put into a superposition of $\ket{0}$ and $\ket{1}$ states with a calibrated microwave pulse (a $\pi/2$ rotation on the Bloch sphere). The spin state then evolves for a time $\tau$ according to the interactions included in the Hamiltonian. A second microwave pulse projects the phase-shifted state back onto $\ket{0}$ and $\ket{1}$ states for readout.

During readout, maximum spin-state contrast is achieved by collecting signal photons during the first 100~ns to 200~ns of green light illumination before the NV center's spin state is eventually reinitialized to $\ket{0}$.  Although Bayesian inference and photon timing data can improve the fidelity,\cite{dinani_bayesian_2019}, signal photons must be collected over many repeats of the sequence.   As shown in fig.\ \ref{fig:figure1}, a count of background photons is collected in a second time window to monitor and compensate for slow changes in the excitation laser intensity.

\subsection{Model}

Each epoch of $m_{\rm s}$ repeated measurement sequences yields a number $n_{\rm s}$ of spin-state-dependent signal photons that depends on the precession time $\tau$.  The goal of the measurement is to determine the time-independent parameters $\theta \equiv \{a, c, \omega_0, T_2\}$ of a ratio model $R(\theta)$:
\begin{eqnarray}
 \lambda_s(t) & = & R(\theta)\lambda_b(t) \\
 R(\theta) & = &  a \left\{ 1 + \frac{c}{2}\left[1 +
 \cos(\omega_0 \tau) \right] e^{-(\tau/T_2)^2}\right\} 
    \label{eq:model_def}
\end{eqnarray}
The counts-per-measurement-sequence signal rate $\lambda_s(t)$ and corresponding background count rate $\lambda_b(t)$ are assumed to be slowly time dependent to account for fluctuations in laser power. The signal photon count $n_{\rm s}$ for an epoch comprising $m_{\rm s}$ repeats of a sequence may be used to calculate an estimate for $\lambda_s(t)$.  Similarly, the background count rate may be estimated from $n_{\rm b}$ background counts over $m_{\rm b}$ sequences.
\begin{eqnarray}
    <n_s(t)> & = & m_s \lambda_s(t), \label{eq:ns_def}\\
    <n_b(t)> & = & m_b \lambda_b(t). \label{eq:nb_def}
\end{eqnarray}
The number of background-counting sequences $m_{\rm b}$ may be chosen independently of $m_{\rm s}$.  We choose $m_{\rm b} \gg m_{\rm s}$ through a moving average, in order to reduce the noise contribution of the background measurements. In order to effectively account for drift, e.g. changes in laser intensity, the time corresponding to $m_{\rm b}$ sequences must be shorter than any characteristic time of the drift.
 
\subsection{Data simulation}
As a guide to measurement simulations, we use the Ramsey experimental data from Ulm University \cite{gentile_experimental_2019} and the descriptions provided in ref.\ \onlinecite{santagati_magnetic-field_2019}. The experimental overhead includes approximately 3~$\mu$s of laser illumination, 1~$\mu$s of relaxation time and 70 ns for microwave pulses and photodetector for a total overhead of 4.07~$\mu$s per sequence. Allowed values of $\tau$ range from 0.1~$\mu$s to 20~$\mu$s of precession time with 50~ns resolution.   

We simulate photon counts using (\ref{eq:model_def}) with ``true'' parameter values $a=0.8$, $c=0.13$, $\omega_0 = 9.4$ $\mu$s$^{-1}$, and either $T_2 = 10$~ns or $T_2 \rightarrow \infty$. Background measurements yield $n_b \approx$ 0.15 photons per sequence on average.  For repeated sequences we use (\ref{eq:ns_def}) and (\ref{eq:nb_def}) with photon counts drawn from Poisson distributions.

\subsection{Experiment Design}

While the goal of the measurement is to estimate the model parameters, the goal of experiment design is to choose $\tau$ values to make the measurements efficiently.  In the simulations, we compare the effectiveness of three protocols: sequential Bayesian experiment design, an adaptive heuristic design, and a non-adaptive design.

Here, we provide a brief overview of sequential Bayesian experiment design. More detailed descriptions may be found in the references.\cite{chaloner_bayesian_1995, granade_robust_2012, granade_qinfer_2017, huan_simulation-based_2013, dushenko_sequential_2020, mcmichael_optbayesexpt_2021}Bayesian experiment design combines Bayesian inference with decision theory to provide an adaptive method for choosing measurement settings. A review by Chaloner and Verdinelli\cite{chaloner_bayesian_1995} describes the development of the methods. ``Unknown'' model parameters are treated as random variables with probability distributions, e.g. with mean and standard deviation corresponding to value and uncertainty. A previous paper demonstrated order-of-magnitude efficiency gains through sequential Bayesian experiment design in an optically detected magnetic resonance experiment.\cite{dushenko_sequential_2020} 

The first task of Bayesian experiment design is to incorporate new measurement data, using Bayesian inference to refine the parameter probability distribution. Bayes' rule yields the {\em posterior} parameter distribution $P(\theta | n_s, m_s )$ given $n_s$ signal photons counted in $m_s$ sequences. The $P(A|B)$ notation is the conditional probability of $A$ given $B$.  The posterior is derived from the {\em prior} distribution $P(\theta)$, and the {\em likelihood} of the measured results 
\begin{equation}
    P(\theta | n_s, m_s ) \propto P(n_s | m_s, \theta, n_b, m_b) P(\theta)
\end{equation}
with likelihood
\begin{equation}
    P(n_s | m_s, \theta, n_b, m_b) \propto  R(\theta)^{n_s}
    \left[\frac{m_s + m_b}{m_s R(\theta) + m_b}\right]^{n_s+n_b}.
    \label{eq:pseudo_likelihood}
\end{equation}
Here, the background rate distribution is estimated from from $n_b$ photons detected in the background channel over $m_b$ repeats, and Poisson-distributed photon counts are assumed. The derivation of this likelihood expression is given in Appendix \ref{app:derivation}.

Once new data has been incorporated in to a refined parameter distribution, the next task is experiment design: choosing settings to make best use of measurement resources. First, the experiment model and measurement noise model are used with the distribution of parameters to predict measurement values for a candidate setting. These predicted values are further used to predict an average improvement of the parameter distribution for that setting. The improvement metric, or {\em utility} targets reduction in the information entropy of the parameter distribution.  Utility calculations are made for all allowed settings, and settings yielding high utility values are selected for future measurements\cite{chaloner_bayesian_1995, granade_robust_2012, huan_simulation-based_2013, dushenko_sequential_2020}.  A non-rigorous interpretation of this method is that high-utility settings tend to be those where the parameter distribution produces a wide distribution of modeled values compared to measurement noise. As a shorthand, we refer to this protocol as ``Bayes.'' Software and documentation is available through ref.\ \onlinecite{mcmichael_optbayesexpt_2021}.

For comparison, we consider an adaptive design for Ramsey experiments reported by Santagati et al.\cite{santagati_magnetic-field_2019}  This heuristic design, ``Tau'' is attractively simple, with the $\tau$ setting determined by $\tau = h/\sigma_{\omega_0}$, where $\sigma_{\omega_0}$ is the width of the frequency distribution and $h \approx 1$ is a tuning parameter. Empirically, we found that a value of $h \approx 0.5$ generated the best results in our simulations.
The Tau protocol shares features with adaptive phase estimation schemes used in Shor's quantum algorithm for prime factorization\cite{shor_algorithms_1994, kitaev_quantum_1996,hayes_swarm_2014}, and a scaling argument (see Appendix \ref{app:TauScaling}) suggests that the Tau protocol might show $1/T$ scaling, but the reported behavior appears to be closer to exponential decay.\cite{santagati_magnetic-field_2019}

Third, we consider a non-adaptive protocol, ``Random,'' where $\tau$ is selected randomly from the allowed settings. This protocol provides a baseline, non-adaptive case for comparison with the adaptive protocols. 

While the three protocols differ in their choices for $\tau$ settings, simulated measurement data were all analyzed using Bayesian inference and the likelihood given by (\ref{eq:pseudo_likelihood}). For all three design protocols, we value elapsed laboratory time, $t_{\rm lab}$ as the resource to be allocated. Accordingly, measurement efficiency is judged on the basis of elapsed time. Therefore, in Tau and Random runs, a consistent time interval is allocated to each epoch, and the Ramsey sequence is repeated with the selected $\tau$ until the time is spent. In the Bayes runs, the measurement time is determined by computation time, as described in the following subsection.

\subsection{Concurrent design and measurement}
\begin{figure}[tb]
    \centering
    \includegraphics[width=\figwidth]{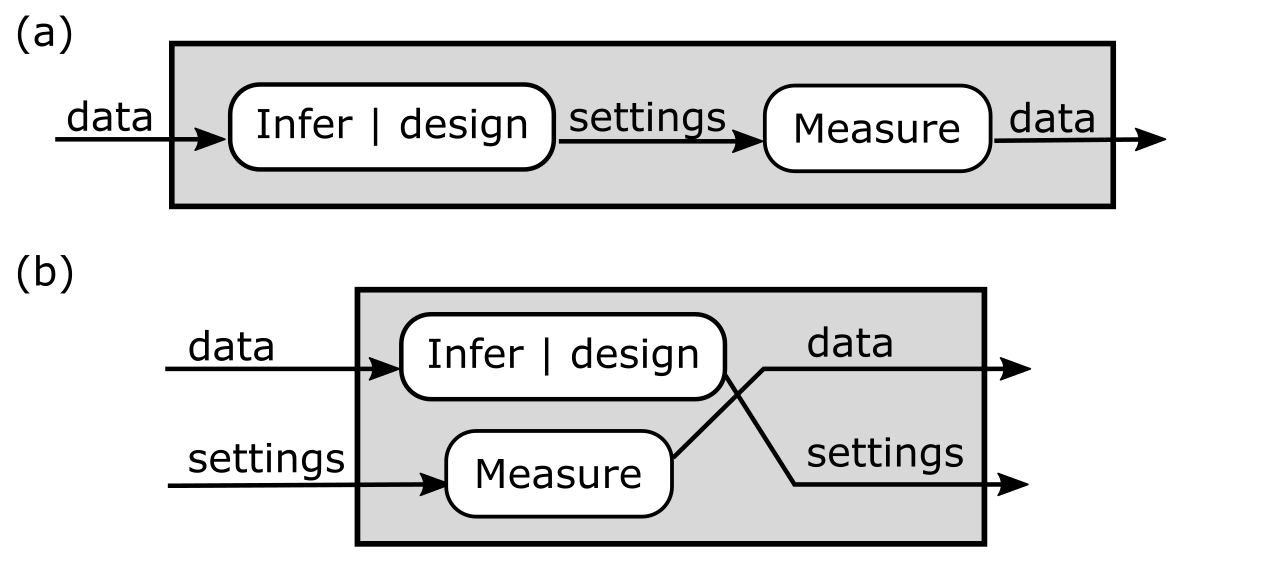}
    \caption{Flow diagrams of computation and measurement using adaptive experiment design. Grey blocks indicate single measurement epochs. In a series workflow (a), data collection stops for data analysis and new setting decisions. In a concurrent workflow (b), measurements continue until new settings are ready.}
    \label{fig:flowcharts}
\end{figure}
The computational time $t_{\rm calc}$ demanded by sequential Bayesian experiment design may be the method's greatest disadvantage to date. In a measure-infer-design loop as illustrated in fig.\ \ref{fig:flowcharts}(a), the measurement waits for instructions during the design calculations. This idle time adds to the experimental overhead and degrades the overall efficiency of the protocol. In the calculations for this paper, computation required a few milliseconds per epoch, which is an enormous overhead compared to a Ramsey $\tau$ of a few microseconds.   

To avoid adding computation time to overhead, we propose a concurrent workflow where measurements and design calculations run at the same time (See fig.\ \ref{fig:flowcharts}(b)).  In epoch $i$, the measurement process accumulates and averages data $y_i$ using a setting $d_{i}$ designed in the previous iteration $i-1$. Meanwhile, Bayesian regression incorporates the previous epoch's measurement data $y_{i-1}$ and chooses settings $d_{i+1}$ for the next iteration.  After the calculations are completed, the measurement system reports data and receives the next setting design. By running measurements and calculations concurrently, the measurement can continue to collect data virtually nonstop. The fact that $d_{i+1}$ is based off accumulated data $\{y_0, \ldots, y_{i-1}\}$, i.e. not including the current measurements $y_i$ has negligible effects in measurements that iterate for many epochs.

We also suggest that for adaptive measurement schemes, the relevant time scale is the time $t_{\rm SNR}$ required to reach a signal to noise ratio (SNR) $\approx$ 1. We propose that the infer/design calculations are fast enough if they can generate designs in a time comparable to the time required for measurement data to significantly change the parameter distribution.  The parameter distribution only changes significantly following measurements with SNR $\gtrsim$ 1 or equivalent.  Using shorter measurement times with fewer repetitions would produce incremental changes in the parameter distribution between computations. At the other extreme, many more repetitions between calculations might hurt efficiency by missing opportunities to switch to higher-utility settings.

In our simulations, $t_{\rm SNR}$ can be estimated. Noise standard deviation equal to the contrast of $c=0.13$ on a background of $a = 0.8$ would require $\approx$ 40 signal photons, or $m_s\approx300$ repeats, or about 4.2 ms at the median setting $\tau = 10$~$\mu$s. By coincidence, in the Bayes runs, the mean measurement time was 4.4 ms, which is comparable to  $t_{\rm SNR}$.  

\section{Results and Discussion\label{sec:results}}

\begin{figure}[tb]
    \centering
    \includegraphics[width=\figwidth]{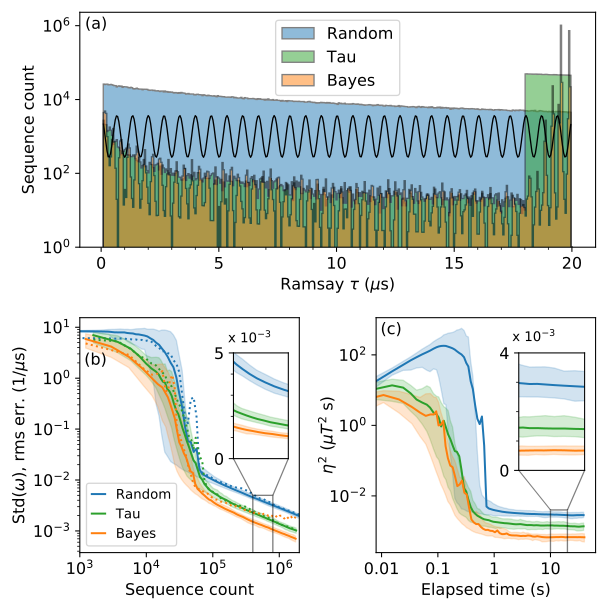}
    \caption{Simulated measurements of $\omega_0$. (a) Average histograms of $\tau$ settings chosen by three protocols: random selection (``Random'', blue), a $\tau = 1/(2\sigma_{\omega_0})$ heuristic (``Tau'', green), and Bayesian experiment design (``Bayes'', orange).  The black sinusoid is the model function evaluated using the true parameter values. (b) Evolution of mean $\sigma_{\omega_0}$. The Bayesian protocol outperforms the heuristic protocol, and the random protocol produces roughly double the uncertainty. Shaded areas are 90 \% credibility regions. Dotted lines are standard deviation of error. (c) Evolution of absolute sensitivity vs. elapsed "wall clock" time. The Bayes protocol is twice as fast as the Tau protocol and roughly four times as fast as the Random protocol. Statistics are calculated over 100 individual runs.}  
    \label{fig:obe_vs_tau}
\end{figure}

Here we compare the performance of Bayesian experiment design with previously investigated methods.  In particular, we test sequential Bayesian experiment design (``Bayes'') against the adaptive heuristic method of Santagati et al.\cite{santagati_magnetic-field_2019} (``Tau'') and random selection of settings (``Random'').

Figure\ \ref{fig:obe_vs_tau} displays statistics from 100-run batches of simulated Ramsey measurements where the precession frequency ${\omega_0}$ is the only unknown parameter. Measurements in Tau and Random runs are allocated 4 ms of measurement time per epoch, i.e. $m_s = 4\;{\rm ms}/(\tau +4.07\; \mu{\rm s})$ repeats of the Ramsey sequence with 4.07~$\mu$s overhead time. Because more repeats are possible for shorter sequences, the Random histogram of sequence counts in fig.\ \ref{fig:obe_vs_tau}(a) appears skewed toward low $\tau$ when measurement time is allocated uniformly.  In Tau runs, $\sigma_{\omega_0}$ eventually becomes small enough that $1/(2\sigma_{\omega_0}) >$ 20~$\mu$s is outside the allowed setting range.  In this case, $\tau$ is selected randomly from the top 10 \% of $\tau$ values. 

Bayes measurement simulations mimic a concurrent workflow, using the execution times of the inference and experiment design code to determine the measurement times. In these simulations, the mean measurement time is 4.4~ms. 

The Tau and Bayes protocols produce similar overall distributions of setting choices, shown in fig.\ \ref{fig:obe_vs_tau}(a).  These protocols both exhibit an initial emphasis on small $\tau$, later moving to large $\tau$. A striking feature of the Bayes histogram of selected $\tau$ values is the comb-like structure indicating that the Bayesian method adaptively concentrates $\tau$ values where the model true curve has maximum slope. By comparison, the Tau and Random protocols show no such phase selectivity, although such selectivity could be programmed into an improved heuristic.

The standard deviation of the $\omega$ distributions are plotted in fig.\ \ref{fig:obe_vs_tau}(b) as a function of the number of Ramsey sequences.  Above $\approx 10^5$ sequences, all three protocols converge to a $(\Sigma m_s)^{-1/2}$ scaling. The inset replots the boxed region on linear scales over a doubling of the sequence count.

Figure \ref{fig:obe_vs_tau}(c) provides a comparison of the absolute sensitivity $\eta$ of the Ramsey ${\omega_0}$ measurement used as a magnetometer. The uncertainties in field and frequency, $\sigma_B$ and $\sigma_{\omega_0}$ respectively, are related by $\sigma_B = \sigma_{\omega_0}/\gamma$, where $\gamma = 2\pi\, 28$~GHz/T is the gyromagnetic ratio. For uncertainties that scale as $t_{\rm lab}^{-1/2}$, $\eta^2 \equiv \sigma_B^2t_{\rm lab}$ is a constant. All three protocols produce nearly constant $\eta^2$ after about 1~s of laboratory time.  The inset replots the boxed region on linear scales. To achieve equivalent uncertainties, the Bayes protocol is about twice as fast as Tau and about four times as fast as Random.

\begin{figure}[tb]
    \centering
    \includegraphics[width=\figwidth]{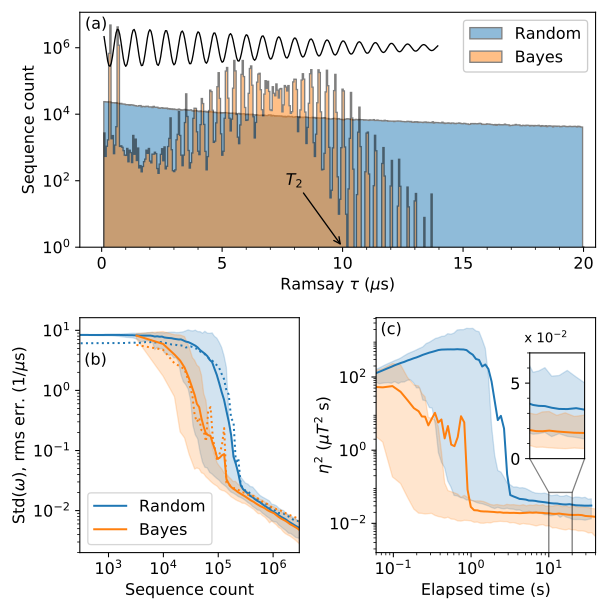}
    \caption{Simulations of measurements to estimate $\omega_0$, $a$, $c$ and $T_2$. (a) Histograms of $\tau$ setting values determined by random selection and by sequential Bayesian experiment design. The black sinusoid is the model function evaluated using true parameter values. (b) Uncertainty in frequency, $\sigma_{\omega_0}$ vs. Ramsey sequence count. (c) Absolute uncertainty $\eta^2$ for Bayesian and random selections of $\tau$. The Bayesian method uses concurrent measurement and calculation. In the random design data, the time required for data analysis has been discounted.}
    \label{fig:seq_vs_concurrent}
\end{figure}

Fig.\ \ref{fig:seq_vs_concurrent} presents results from 100-run batches of measurement simulations where $a$, $c$, $\omega_0$, and $T_2$ are all treated as unknowns.  We compare only Bayes and Random protocols, as no heuristic is available. As above, 4 ms were allotted for measurements in the Random protocol, but Ramsey code execution was slower with 4 variables, and an average of 13 ms were allocated per epoch in Bayes protocol runs.  

Compared to the histogram in fig.\ \ref{fig:obe_vs_tau}(a), the histogram structure in fig.\ \ref{fig:seq_vs_concurrent}(a) is much richer. For a qualitative interpretation of this structure, we note that the modeled signal given by (\ref{eq:model_def}) is most sensitive to contrast $c$ at small $\tau$ extrema, is most sensitive to $T_2$ for extrema near $\tau \approx T_2$( = 10 $\mu$s) and is most sensitive to $\omega_0$ for large slope of the model function and $\tau \approx T_2 / \sqrt{2}$. The maxima of the envelope are skewed to slightly smaller $\tau$ values by the fact that low-$\tau$ designs produce more photons per epoch, and deliver smaller uncertainty.

Presumably, an adaptive heuristic could be developed and tuned for this four-variable measurement. It is likely that the development would require significant labor, however. Implementation of the Bayes protocol was comparatively simple, requiring only minor adjustments to include $a$, $c$ and $T_2$ as unknowns in the model function.

Figs.\ \ref{fig:seq_vs_concurrent}(b)-(c) show that the Random protocol is slower to converge than the Bayes protocol, but the contrast in performance is weaker than in the single-unknown case above. Fig.\ \ref{fig:seq_vs_concurrent}(b) shows that the Random and Bayes protocols achieve equivalent uncertainty levels for equal number of sequences run, or photons collected. However, fig.\ \ref{fig:seq_vs_concurrent}(c) shows that the Bayes protocol makes more efficient use of laboratory time.  We attribute the apparent efficiency of the random approach to the fact high-utility $\tau$ values are widely distributed among the available settings, not tightly grouped at large $\tau$ as they appear in fig.\ \ref{fig:obe_vs_tau}(a).

\section{Summary}

The simulation results show that sequential Bayesian experiment design is an efficient protocol for single-NV Ramsey measurements, outperforming a tuned heuristic ``Tau'' protocol and random setting selection in efficient use of laboratory, ``wall-clock''time. A key factor for the Bayes protocol is the introduction of a concurrent workflow, which allows measurements to continue until design calculations are complete, and effectively eliminates computation time from the overhead, at least when computation time is less than the time needed to attain SNR $\approx 1$.  Longer measurements might reduce efficiency through missed opportunities to select higher-utility settings.

The efficiency gains produced by sequential Bayesian experiment design in this work are significant, but modest compared to the order-of-magnitude gains previously reported for measurements of peaks on a wide, flat background.\cite{dushenko_sequential_2020} We explain this difference noting that measurements in the broad background of a spectrum offer little compared to the information-rich settings near the peaks, so large gains are possible by focusing measurements on the peak regions.  In contrast, the Ramsey measurements tend to be informative over a larger range of setting values. 

While this paper has focused on efficiency during data acquisition, sequential Bayesian experiment design also offers efficiency advantages in the time periods before and after data acquisition. For development of a protocol before measurement, the Bayes methods are easily adapted to new experiment models or to different measurement goals, especially when compared to the demands of designing efficient heuristics. Even in cases where Bayes calculations would be prohibitively slow, the Bayes protocols may serve as a reliable guide for heuristic protocol development.
Also, if post-measurement data analysis is required, it might be reasonable to count analysis time as part of measurement overhead. In the Random protocol results above, data was analyzed at each epoch to show progress but the analysis time was discounted. But in a more typical use, a Random protocol would yield raw data, and any required data analysis might contribute to overhead.  The Bayes protocol offers (almost) instant results because the data is continuously analyzed as part of the protocol.

The data that support the findings of this study are available from the corresponding author upon reasonable request.

\begin{acknowledgments}
S.D. acknowledges support under the cooperative research agreement between the University of Maryland and the National Institute of Standards and Technology Physical Measurement Laboratory (Grant No. 70NANB14H209) through the University of Maryland. The authors thank Adam Pintar for many helpful discussions.
\end{acknowledgments}

\appendix

\section{Derivation of likelihood\label{app:derivation}}

This appendix provides a derivation of the likelihood $P(n_s, m_s, |n_b, m_b \theta)$ of receiving $n_s$ signal counts in $m_s$ repeats, given $n_b$ background counts in $m_b$ repeats such that
\begin{equation}
    P(\theta|n_s, m_s, n_b, m_b) = P(n_s | m_s, n_b, m_b, \theta)P(\theta),
\end{equation}
where $P(\theta)$ incorporates $n$ and $m$ data from all preceding epochs.
The derivation expresses the likelihood of $n_b$ counts in $m_b$ repeats as a function of a signal count rate $\lambda_s$, which is the product of the ratio $R(\theta)$ and the background count $\lambda_b$. The distribution $P(\lambda_b)$ is determined from $n_b$ and $m_b$, then eliminated from the final expression by integration. 

The likelihood of receiving $n_s$ signal counts in $m_s$ repeats is a Poisson distribution with mean $m_s R(\theta)\lambda_b$.
\begin{equation}
    P(n_s | m_s, \theta, \lambda_b) =
    \frac{(m_s R(\theta)\lambda_b)^{n_s}e^{-m_s R(\theta)\lambda_b}}{n_s!}.
    \label{eq:ns_likelihood}
\end{equation}
The background count rate distribution is determined by $n_b$ and $n_b$
\begin{equation}
    P(\lambda_b) \propto \frac{(m_b\lambda_b)^{n_b}e^{-m_b\lambda_b}}{n_b!}
      \lambda_b^\nu.
      \label{eq:Plambdab}
\end{equation}
The fraction in this expression is a Poisson distribution expressing the likelihood of counting $n_b$ photons, and the trailing $\lambda_b^\nu$ is a prior with exponent $\nu$ to be determined later. For the Jeffreys prior, $\nu = -1/2$.

The distribution of $\lambda_b$ values is incorporated by integration.
\begin{equation}
    P(n_s| m_s, n_b, m_b, \theta) = \int P(n_s | m_s, \theta, \lambda_b) P(\lambda_b | n_b, m_b) d\lambda_b
\end{equation}
Substituting from (\ref{eq:ns_likelihood}) and (\ref{eq:Plambdab}), the integral is tractable, yielding
\begin{equation}
    P(n_s| m_s, n_b, m_b, \theta) = C \frac{R(\theta)^{n_s}}
    {\left[m_s R(\theta) + m_b\right]^{n_s + n_b + 1 + \nu}}.
\end{equation}

The prior exponent $\nu$ is selected so that the net effect of two likelihoods from two identical measurement results is the same as a single likelihood with combined data. That is, we require
\begin{equation}
    P(n_s| m_s, n_b, m_b, \theta)^2 \propto P(2n_s| 2m_s, 2n_b, 2m_b, \theta),
\end{equation}
which is satisfied by $\nu = -1$.  In tests of the likelihood using $\nu = 0$, the simulations converged to incorrect mean parameter values. With $\nu = -1$ the simulations regularly converged to the true values within a few standard deviations. We also note that the choice of $\nu = -1$ a gamma distribution prior of the general form.
\begin{equation}
    \lambda_b \sim \Gamma(\alpha, \beta) 
    = \frac{\beta^{\alpha} \lambda_b^{\alpha -1} e^{-\beta \lambda_b}}
    {\Gamma(\alpha)},
\end{equation}
where $\Gamma(\cdot)$ is the gamma function. Gamma distributions are conjugate priors for Poisson likelihoods, and in the case $\alpha \rightarrow 0$ and $\beta \rightarrow 0$, the gamma distribution becomes ``vague'' as variance $\rightarrow \infty$.

Since the $\theta$ distribution is explicitly normalized in software, we are free to choose a convenient pseudo normalization for convenience. The factor $C$ contains exponential functions $m_s!$, $m_n!$, and $(m_s + m_b)!$, which are challenging for large $m$ values.  To eliminate factorials we choose the following normalization.
\begin{equation}
        P(n_s | m_s, \theta, n_b, m_b) \propto  R(\theta)^{n_s}
    \left[\frac{m_s + m_b}{m_s R(\theta) + m_b}\right]^{n_s+n_b}.
    \label{eq:likelihood_in_appendix}
\end{equation}

\begin{figure}[tb]
    \centering
    \includegraphics[width=\figwidth]{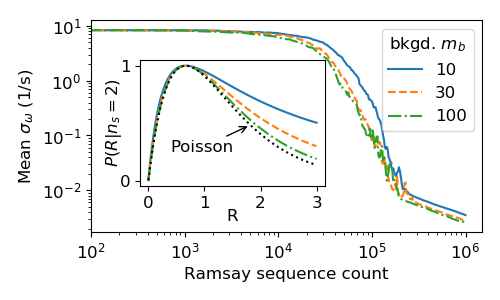}
    \caption{Influence of background count averaging on the likelihood (inset) and on the performance of the Bayesian protocol. For the inset, $m_s = 10$, $n_s = 1$, $n_b = 0.15 m_s$, and the true $R = 0.1/0.15$.  In both the main figure an inset, the benefits of background averaging saturate for $n_b \gtrsim 10 \times n_s$. The line labeled ``Poisson'' is the $R$ distribution assuming $\lambda_b = 0.15$.}
    \label{fig:likelihood_demo}
\end{figure}

Fig. \ref{fig:likelihood_demo} illustrates the behavior of (\ref{eq:pseudo_likelihood}) or (\ref{eq:likelihood_in_appendix}) as a function of the number of background measurements, $m_b$, included  in the background count $n_b$. The inset shows that the likelihood is a peaked function that approaches a Poisson distribution as $m_b$ increases.  The main plot shows the effect of the $m_s$ on the performance of the Bayes method. Only marginal improvements are gained by extending background measurements beyond $n_b \gtrsim 10 \times n_s$.

\section{Scaling of Tau protocol\label{app:TauScaling}}

Each epoch begins with a consistent phase uncertainty $\sigma_\phi = \tau\sigma_{\omega_0} = h$. If the number of repeats per epoch is constant, each epoch decreases uncertainty by a constant factor $\beta < 1$ on average. After the $k^{\rm th}$ epoch,
\begin{equation}
    \sigma \propto \beta^k
\end{equation}
Each epoch requires measurement time proportional to $\tau$, neglecting overhead, so total measurement time $T$ scales as 
\begin{equation}
    T \propto \beta^{-(k+1)}
\end{equation}
The net behavior is therefore predicted to follow Heisenberg scaling,
\begin{equation}
    \sigma \propto 1/T,
\end{equation}
under the assumption that the number of repeats per epoch is constant.

\newpage

\bibliographystyle{aipnum4-1}
\bibliography{Ramsey}

\begin{thebibliography}{59}%
\makeatletter
\providecommand \@ifxundefined [1]{%
 \@ifx{#1\undefined}
}%
\providecommand \@ifnum [1]{%
 \ifnum #1\expandafter \@firstoftwo
 \else \expandafter \@secondoftwo
 \fi
}%
\providecommand \@ifx [1]{%
 \ifx #1\expandafter \@firstoftwo
 \else \expandafter \@secondoftwo
 \fi
}%
\providecommand \natexlab [1]{#1}%
\providecommand \enquote  [1]{``#1''}%
\providecommand \bibnamefont  [1]{#1}%
\providecommand \bibfnamefont [1]{#1}%
\providecommand \citenamefont [1]{#1}%
\providecommand \href@noop [0]{\@secondoftwo}%
\providecommand \href [0]{\begingroup \@sanitize@url \@href}%
\providecommand \@href[1]{\@@startlink{#1}\@@href}%
\providecommand \@@href[1]{\endgroup#1\@@endlink}%
\providecommand \@sanitize@url [0]{\catcode `\\12\catcode `\$12\catcode
  `\&12\catcode `\#12\catcode `\^12\catcode `\_12\catcode `\%12\relax}%
\providecommand \@@startlink[1]{}%
\providecommand \@@endlink[0]{}%
\providecommand \url  [0]{\begingroup\@sanitize@url \@url }%
\providecommand \@url [1]{\endgroup\@href {#1}{\urlprefix }}%
\providecommand \urlprefix  [0]{URL }%
\providecommand \Eprint [0]{\href }%
\providecommand \doibase [0]{http://dx.doi.org/}%
\providecommand \selectlanguage [0]{\@gobble}%
\providecommand \bibinfo  [0]{\@secondoftwo}%
\providecommand \bibfield  [0]{\@secondoftwo}%
\providecommand \translation [1]{[#1]}%
\providecommand \BibitemOpen [0]{}%
\providecommand \bibitemStop [0]{}%
\providecommand \bibitemNoStop [0]{.\EOS\space}%
\providecommand \EOS [0]{\spacefactor3000\relax}%
\providecommand \BibitemShut  [1]{\csname bibitem#1\endcsname}%
\let\auto@bib@innerbib\@empty
\bibitem [{\citenamefont {Chernobrod}\ and\ \citenamefont
  {Berman}(2005)}]{chernobrod_spin_2005}%
  \BibitemOpen
  \bibfield  {author} {\bibinfo {author} {\bibfnamefont {B.~M.}\ \bibnamefont
  {Chernobrod}}\ and\ \bibinfo {author} {\bibfnamefont {G.~P.}\ \bibnamefont
  {Berman}},\ }\href
  {http://scitation.aip.org/content/aip/journal/jap/97/1/10.1063/1.1829373}
  {\bibfield  {journal} {\bibinfo  {journal} {Journal of applied physics}\
  }\textbf {\bibinfo {volume} {97}},\ \bibinfo {pages} {014903} (\bibinfo
  {year} {2005})}\BibitemShut {NoStop}%
\bibitem [{\citenamefont {Balasubramanian}\ \emph {et~al.}(2008)\citenamefont
  {Balasubramanian}, \citenamefont {Chan}, \citenamefont {Kolesov},
  \citenamefont {Al-Hmoud}, \citenamefont {Tisler}, \citenamefont {Shin},
  \citenamefont {Kim}, \citenamefont {Wojcik}, \citenamefont {Hemmer},
  \citenamefont {Krueger}, \citenamefont {Hanke}, \citenamefont
  {Leitenstorfer}, \citenamefont {Bratschitsch}, \citenamefont {Jelezko},\ and\
  \citenamefont {Wrachtrup}}]{balasubramanian_nanoscale_2008}%
  \BibitemOpen
  \bibfield  {author} {\bibinfo {author} {\bibfnamefont {G.}~\bibnamefont
  {Balasubramanian}}, \bibinfo {author} {\bibfnamefont {I.~Y.}\ \bibnamefont
  {Chan}}, \bibinfo {author} {\bibfnamefont {R.}~\bibnamefont {Kolesov}},
  \bibinfo {author} {\bibfnamefont {M.}~\bibnamefont {Al-Hmoud}}, \bibinfo
  {author} {\bibfnamefont {J.}~\bibnamefont {Tisler}}, \bibinfo {author}
  {\bibfnamefont {C.}~\bibnamefont {Shin}}, \bibinfo {author} {\bibfnamefont
  {C.}~\bibnamefont {Kim}}, \bibinfo {author} {\bibfnamefont {A.}~\bibnamefont
  {Wojcik}}, \bibinfo {author} {\bibfnamefont {P.~R.}\ \bibnamefont {Hemmer}},
  \bibinfo {author} {\bibfnamefont {A.}~\bibnamefont {Krueger}}, \bibinfo
  {author} {\bibfnamefont {T.}~\bibnamefont {Hanke}}, \bibinfo {author}
  {\bibfnamefont {A.}~\bibnamefont {Leitenstorfer}}, \bibinfo {author}
  {\bibfnamefont {R.}~\bibnamefont {Bratschitsch}}, \bibinfo {author}
  {\bibfnamefont {F.}~\bibnamefont {Jelezko}}, \ and\ \bibinfo {author}
  {\bibfnamefont {J.}~\bibnamefont {Wrachtrup}},\ }\href {\doibase
  10.1038/nature07278} {\bibfield  {journal} {\bibinfo  {journal} {Nature}\
  }\textbf {\bibinfo {volume} {455}},\ \bibinfo {pages} {648} (\bibinfo {year}
  {2008})}\BibitemShut {NoStop}%
\bibitem [{\citenamefont {Maze}\ \emph {et~al.}(2008)\citenamefont {Maze},
  \citenamefont {Stanwix}, \citenamefont {Hodges}, \citenamefont {Hong},
  \citenamefont {Taylor}, \citenamefont {Cappellaro}, \citenamefont {Jiang},
  \citenamefont {Dutt}, \citenamefont {Togan}, \citenamefont {Zibrov},
  \citenamefont {Yacoby}, \citenamefont {Walsworth},\ and\ \citenamefont
  {Lukin}}]{maze_nanoscale_2008}%
  \BibitemOpen
  \bibfield  {author} {\bibinfo {author} {\bibfnamefont {J.~R.}\ \bibnamefont
  {Maze}}, \bibinfo {author} {\bibfnamefont {P.~L.}\ \bibnamefont {Stanwix}},
  \bibinfo {author} {\bibfnamefont {J.~S.}\ \bibnamefont {Hodges}}, \bibinfo
  {author} {\bibfnamefont {S.}~\bibnamefont {Hong}}, \bibinfo {author}
  {\bibfnamefont {J.~M.}\ \bibnamefont {Taylor}}, \bibinfo {author}
  {\bibfnamefont {P.}~\bibnamefont {Cappellaro}}, \bibinfo {author}
  {\bibfnamefont {L.}~\bibnamefont {Jiang}}, \bibinfo {author} {\bibfnamefont
  {M.~V.~G.}\ \bibnamefont {Dutt}}, \bibinfo {author} {\bibfnamefont
  {E.}~\bibnamefont {Togan}}, \bibinfo {author} {\bibfnamefont {A.~S.}\
  \bibnamefont {Zibrov}}, \bibinfo {author} {\bibfnamefont {A.}~\bibnamefont
  {Yacoby}}, \bibinfo {author} {\bibfnamefont {R.~L.}\ \bibnamefont
  {Walsworth}}, \ and\ \bibinfo {author} {\bibfnamefont {M.~D.}\ \bibnamefont
  {Lukin}},\ }\href {\doibase 10.1038/nature07279} {\bibfield  {journal}
  {\bibinfo  {journal} {Nature}\ }\textbf {\bibinfo {volume} {455}},\ \bibinfo
  {pages} {644} (\bibinfo {year} {2008})}\BibitemShut {NoStop}%
\bibitem [{\citenamefont {Taylor}\ \emph {et~al.}(2008)\citenamefont {Taylor},
  \citenamefont {Cappellaro}, \citenamefont {Childress}, \citenamefont {Jiang},
  \citenamefont {Budker}, \citenamefont {Hemmer}, \citenamefont {Yacoby},
  \citenamefont {Walsworth},\ and\ \citenamefont
  {Lukin}}]{taylor_high-sensitivity_2008}%
  \BibitemOpen
  \bibfield  {author} {\bibinfo {author} {\bibfnamefont {J.~M.}\ \bibnamefont
  {Taylor}}, \bibinfo {author} {\bibfnamefont {P.}~\bibnamefont {Cappellaro}},
  \bibinfo {author} {\bibfnamefont {L.}~\bibnamefont {Childress}}, \bibinfo
  {author} {\bibfnamefont {L.}~\bibnamefont {Jiang}}, \bibinfo {author}
  {\bibfnamefont {D.}~\bibnamefont {Budker}}, \bibinfo {author} {\bibfnamefont
  {P.~R.}\ \bibnamefont {Hemmer}}, \bibinfo {author} {\bibfnamefont
  {A.}~\bibnamefont {Yacoby}}, \bibinfo {author} {\bibfnamefont
  {R.}~\bibnamefont {Walsworth}}, \ and\ \bibinfo {author} {\bibfnamefont
  {M.~D.}\ \bibnamefont {Lukin}},\ }\href {\doibase 10.1038/nphys1075}
  {\bibfield  {journal} {\bibinfo  {journal} {Nature Physics}\ }\textbf
  {\bibinfo {volume} {4}},\ \bibinfo {pages} {810} (\bibinfo {year}
  {2008})}\BibitemShut {NoStop}%
\bibitem [{\citenamefont {Rondin}\ \emph {et~al.}(2014)\citenamefont {Rondin},
  \citenamefont {Tetienne}, \citenamefont {Hingant}, \citenamefont {Roch},
  \citenamefont {Maletinsky},\ and\ \citenamefont
  {Jacques}}]{rondin_magnetometry_2014}%
  \BibitemOpen
  \bibfield  {author} {\bibinfo {author} {\bibfnamefont {L.}~\bibnamefont
  {Rondin}}, \bibinfo {author} {\bibfnamefont {J.-P.}\ \bibnamefont
  {Tetienne}}, \bibinfo {author} {\bibfnamefont {T.}~\bibnamefont {Hingant}},
  \bibinfo {author} {\bibfnamefont {J.-F.}\ \bibnamefont {Roch}}, \bibinfo
  {author} {\bibfnamefont {P.}~\bibnamefont {Maletinsky}}, \ and\ \bibinfo
  {author} {\bibfnamefont {V.}~\bibnamefont {Jacques}},\ }\href {\doibase
  10.1088/0034-4885/77/5/056503} {\bibfield  {journal} {\bibinfo  {journal}
  {Reports on Progress in Physics}\ }\textbf {\bibinfo {volume} {77}},\
  \bibinfo {pages} {056503} (\bibinfo {year} {2014})}\BibitemShut {NoStop}%
\bibitem [{\citenamefont {Hingant}\ \emph {et~al.}(2015)\citenamefont
  {Hingant}, \citenamefont {Tetienne}, \citenamefont {Martínez}, \citenamefont
  {Garcia}, \citenamefont {Ravelosona}, \citenamefont {Roch},\ and\
  \citenamefont {Jacques}}]{hingant_measuring_2015}%
  \BibitemOpen
  \bibfield  {author} {\bibinfo {author} {\bibfnamefont {T.}~\bibnamefont
  {Hingant}}, \bibinfo {author} {\bibfnamefont {J.-P.}\ \bibnamefont
  {Tetienne}}, \bibinfo {author} {\bibfnamefont {L.~J.}\ \bibnamefont
  {Martínez}}, \bibinfo {author} {\bibfnamefont {K.}~\bibnamefont {Garcia}},
  \bibinfo {author} {\bibfnamefont {D.}~\bibnamefont {Ravelosona}}, \bibinfo
  {author} {\bibfnamefont {J.-F.}\ \bibnamefont {Roch}}, \ and\ \bibinfo
  {author} {\bibfnamefont {V.}~\bibnamefont {Jacques}},\ }\href
  {http://journals.aps.org/prapplied/abstract/10.1103/PhysRevApplied.4.014003}
  {\bibfield  {journal} {\bibinfo  {journal} {Physical Review Applied}\
  }\textbf {\bibinfo {volume} {4}},\ \bibinfo {pages} {014003} (\bibinfo {year}
  {2015})}\BibitemShut {NoStop}%
\bibitem [{\citenamefont {Van~der Sar}\ \emph {et~al.}(2014)\citenamefont
  {Van~der Sar}, \citenamefont {Casola}, \citenamefont {Walsworth},\ and\
  \citenamefont {Yacoby}}]{van_der_sar_nanometre-scale_2014}%
  \BibitemOpen
  \bibfield  {author} {\bibinfo {author} {\bibfnamefont {T.}~\bibnamefont
  {Van~der Sar}}, \bibinfo {author} {\bibfnamefont {F.}~\bibnamefont {Casola}},
  \bibinfo {author} {\bibfnamefont {R.}~\bibnamefont {Walsworth}}, \ and\
  \bibinfo {author} {\bibfnamefont {A.}~\bibnamefont {Yacoby}},\ }\href
  {http://arxiv.org/abs/1410.6423} {\bibfield  {journal} {\bibinfo  {journal}
  {arXiv preprint arXiv:1410.6423}\ } (\bibinfo {year} {2014})}\BibitemShut
  {NoStop}%
\bibitem [{\citenamefont {Tetienne}\ \emph {et~al.}(2014)\citenamefont
  {Tetienne}, \citenamefont {Hingant}, \citenamefont {Kim}, \citenamefont
  {Diez}, \citenamefont {Adam}, \citenamefont {Garcia}, \citenamefont {Roch},
  \citenamefont {Rohart}, \citenamefont {Thiaville}, \citenamefont
  {Ravelosona},\ and\ \citenamefont {Jacques}}]{tetienne_nanoscale_2014}%
  \BibitemOpen
  \bibfield  {author} {\bibinfo {author} {\bibfnamefont {J.-P.}\ \bibnamefont
  {Tetienne}}, \bibinfo {author} {\bibfnamefont {T.}~\bibnamefont {Hingant}},
  \bibinfo {author} {\bibfnamefont {J.-V.}\ \bibnamefont {Kim}}, \bibinfo
  {author} {\bibfnamefont {L.~H.}\ \bibnamefont {Diez}}, \bibinfo {author}
  {\bibfnamefont {J.-P.}\ \bibnamefont {Adam}}, \bibinfo {author}
  {\bibfnamefont {K.}~\bibnamefont {Garcia}}, \bibinfo {author} {\bibfnamefont
  {J.-F.}\ \bibnamefont {Roch}}, \bibinfo {author} {\bibfnamefont
  {S.}~\bibnamefont {Rohart}}, \bibinfo {author} {\bibfnamefont
  {A.}~\bibnamefont {Thiaville}}, \bibinfo {author} {\bibfnamefont
  {D.}~\bibnamefont {Ravelosona}}, \ and\ \bibinfo {author} {\bibfnamefont
  {V.}~\bibnamefont {Jacques}},\ }\href {\doibase 10.1126/science.1250113}
  {\bibfield  {journal} {\bibinfo  {journal} {Science}\ }\textbf {\bibinfo
  {volume} {344}},\ \bibinfo {pages} {1366} (\bibinfo {year}
  {2014})}\BibitemShut {NoStop}%
\bibitem [{\citenamefont {Tetienne}\ \emph {et~al.}(2017)\citenamefont
  {Tetienne}, \citenamefont {Dontschuk}, \citenamefont {Broadway},
  \citenamefont {Stacey}, \citenamefont {Simpson},\ and\ \citenamefont
  {Hollenberg}}]{tetienne_quantum_2017}%
  \BibitemOpen
  \bibfield  {author} {\bibinfo {author} {\bibfnamefont {J.-P.}\ \bibnamefont
  {Tetienne}}, \bibinfo {author} {\bibfnamefont {N.}~\bibnamefont {Dontschuk}},
  \bibinfo {author} {\bibfnamefont {D.~A.}\ \bibnamefont {Broadway}}, \bibinfo
  {author} {\bibfnamefont {A.}~\bibnamefont {Stacey}}, \bibinfo {author}
  {\bibfnamefont {D.~A.}\ \bibnamefont {Simpson}}, \ and\ \bibinfo {author}
  {\bibfnamefont {L.~C.~L.}\ \bibnamefont {Hollenberg}},\ }\href {\doibase
  10.1126/sciadv.1602429} {\bibfield  {journal} {\bibinfo  {journal} {Science
  Advances}\ }\textbf {\bibinfo {volume} {3}},\ \bibinfo {pages} {e1602429}
  (\bibinfo {year} {2017})}\BibitemShut {NoStop}%
\bibitem [{\citenamefont {Rondin}\ \emph {et~al.}(2013)\citenamefont {Rondin},
  \citenamefont {Tetienne}, \citenamefont {Rohart}, \citenamefont {Thiaville},
  \citenamefont {Hingant}, \citenamefont {Spinicelli}, \citenamefont {Roch},\
  and\ \citenamefont {Jacques}}]{rondin_stray-field_2013}%
  \BibitemOpen
  \bibfield  {author} {\bibinfo {author} {\bibfnamefont {L.}~\bibnamefont
  {Rondin}}, \bibinfo {author} {\bibfnamefont {J.~P.}\ \bibnamefont
  {Tetienne}}, \bibinfo {author} {\bibfnamefont {S.}~\bibnamefont {Rohart}},
  \bibinfo {author} {\bibfnamefont {A.}~\bibnamefont {Thiaville}}, \bibinfo
  {author} {\bibfnamefont {T.}~\bibnamefont {Hingant}}, \bibinfo {author}
  {\bibfnamefont {P.}~\bibnamefont {Spinicelli}}, \bibinfo {author}
  {\bibfnamefont {J.~F.}\ \bibnamefont {Roch}}, \ and\ \bibinfo {author}
  {\bibfnamefont {V.}~\bibnamefont {Jacques}},\ }\href {\doibase
  10.1038/ncomms3279} {\bibfield  {journal} {\bibinfo  {journal} {Nature
  Communications}\ }\textbf {\bibinfo {volume} {4}},\ \bibinfo {pages} {2279}
  (\bibinfo {year} {2013})}\BibitemShut {NoStop}%
\bibitem [{\citenamefont {Turner}\ \emph {et~al.}(2020)\citenamefont {Turner},
  \citenamefont {Langellier}, \citenamefont {Bainbridge}, \citenamefont
  {Walters}, \citenamefont {Meesala}, \citenamefont {Babinec}, \citenamefont
  {Kehayias}, \citenamefont {Yacoby}, \citenamefont {Hu}, \citenamefont
  {Lončar}, \citenamefont {Walsworth},\ and\ \citenamefont
  {Levine}}]{turner_magnetic_2020}%
  \BibitemOpen
  \bibfield  {author} {\bibinfo {author} {\bibfnamefont {M.~J.}\ \bibnamefont
  {Turner}}, \bibinfo {author} {\bibfnamefont {N.}~\bibnamefont {Langellier}},
  \bibinfo {author} {\bibfnamefont {R.}~\bibnamefont {Bainbridge}}, \bibinfo
  {author} {\bibfnamefont {D.}~\bibnamefont {Walters}}, \bibinfo {author}
  {\bibfnamefont {S.}~\bibnamefont {Meesala}}, \bibinfo {author} {\bibfnamefont
  {T.~M.}\ \bibnamefont {Babinec}}, \bibinfo {author} {\bibfnamefont
  {P.}~\bibnamefont {Kehayias}}, \bibinfo {author} {\bibfnamefont
  {A.}~\bibnamefont {Yacoby}}, \bibinfo {author} {\bibfnamefont
  {E.}~\bibnamefont {Hu}}, \bibinfo {author} {\bibfnamefont {M.}~\bibnamefont
  {Lončar}}, \bibinfo {author} {\bibfnamefont {R.~L.}\ \bibnamefont
  {Walsworth}}, \ and\ \bibinfo {author} {\bibfnamefont {E.~V.}\ \bibnamefont
  {Levine}},\ }\href {\doibase 10.1103/PhysRevApplied.14.014097} {\bibfield
  {journal} {\bibinfo  {journal} {Physical Review Applied}\ }\textbf {\bibinfo
  {volume} {14}},\ \bibinfo {pages} {014097} (\bibinfo {year}
  {2020})}\BibitemShut {NoStop}%
\bibitem [{\citenamefont {Horsley}\ \emph {et~al.}(2018)\citenamefont
  {Horsley}, \citenamefont {Appel}, \citenamefont {Wolters}, \citenamefont
  {Achard}, \citenamefont {Tallaire}, \citenamefont {Maletinsky},\ and\
  \citenamefont {Treutlein}}]{horsley_microwave_2018}%
  \BibitemOpen
  \bibfield  {author} {\bibinfo {author} {\bibfnamefont {A.}~\bibnamefont
  {Horsley}}, \bibinfo {author} {\bibfnamefont {P.}~\bibnamefont {Appel}},
  \bibinfo {author} {\bibfnamefont {J.}~\bibnamefont {Wolters}}, \bibinfo
  {author} {\bibfnamefont {J.}~\bibnamefont {Achard}}, \bibinfo {author}
  {\bibfnamefont {A.}~\bibnamefont {Tallaire}}, \bibinfo {author}
  {\bibfnamefont {P.}~\bibnamefont {Maletinsky}}, \ and\ \bibinfo {author}
  {\bibfnamefont {P.}~\bibnamefont {Treutlein}},\ }\href {\doibase
  10.1103/PhysRevApplied.10.044039} {\bibfield  {journal} {\bibinfo  {journal}
  {Physical Review Applied}\ }\textbf {\bibinfo {volume} {10}},\ \bibinfo
  {pages} {044039} (\bibinfo {year} {2018})}\BibitemShut {NoStop}%
\bibitem [{\citenamefont {Appel}\ \emph {et~al.}(2015)\citenamefont {Appel},
  \citenamefont {Ganzhorn}, \citenamefont {Neu},\ and\ \citenamefont
  {Maletinsky}}]{appel_nanoscale_2015}%
  \BibitemOpen
  \bibfield  {author} {\bibinfo {author} {\bibfnamefont {P.}~\bibnamefont
  {Appel}}, \bibinfo {author} {\bibfnamefont {M.}~\bibnamefont {Ganzhorn}},
  \bibinfo {author} {\bibfnamefont {E.}~\bibnamefont {Neu}}, \ and\ \bibinfo
  {author} {\bibfnamefont {P.}~\bibnamefont {Maletinsky}},\ }\href {\doibase
  10.1088/1367-2630/17/11/112001} {\bibfield  {journal} {\bibinfo  {journal}
  {New Journal of Physics}\ }\textbf {\bibinfo {volume} {17}},\ \bibinfo
  {pages} {112001} (\bibinfo {year} {2015})}\BibitemShut {NoStop}%
\bibitem [{\citenamefont {Kucsko}\ \emph {et~al.}(2013)\citenamefont {Kucsko},
  \citenamefont {Maurer}, \citenamefont {Yao}, \citenamefont {Kubo},
  \citenamefont {Noh}, \citenamefont {Lo}, \citenamefont {Park},\ and\
  \citenamefont {Lukin}}]{kucsko_nanometre-scale_2013}%
  \BibitemOpen
  \bibfield  {author} {\bibinfo {author} {\bibfnamefont {G.}~\bibnamefont
  {Kucsko}}, \bibinfo {author} {\bibfnamefont {P.~C.}\ \bibnamefont {Maurer}},
  \bibinfo {author} {\bibfnamefont {N.~Y.}\ \bibnamefont {Yao}}, \bibinfo
  {author} {\bibfnamefont {M.}~\bibnamefont {Kubo}}, \bibinfo {author}
  {\bibfnamefont {H.~J.}\ \bibnamefont {Noh}}, \bibinfo {author} {\bibfnamefont
  {P.~K.}\ \bibnamefont {Lo}}, \bibinfo {author} {\bibfnamefont
  {H.}~\bibnamefont {Park}}, \ and\ \bibinfo {author} {\bibfnamefont {M.~D.}\
  \bibnamefont {Lukin}},\ }\href {\doibase 10.1038/nature12373} {\bibfield
  {journal} {\bibinfo  {journal} {Nature}\ }\textbf {\bibinfo {volume} {500}},\
  \bibinfo {pages} {54} (\bibinfo {year} {2013})}\BibitemShut {NoStop}%
\bibitem [{\citenamefont {McGuinness}\ \emph {et~al.}(2011)\citenamefont
  {McGuinness}, \citenamefont {Yan}, \citenamefont {Stacey}, \citenamefont
  {Simpson}, \citenamefont {Hall}, \citenamefont {Maclaurin}, \citenamefont
  {Prawer}, \citenamefont {Mulvaney}, \citenamefont {Wrachtrup}, \citenamefont
  {Caruso}, \citenamefont {Scholten},\ and\ \citenamefont
  {Hollenberg}}]{mcguinness_quantum_2011}%
  \BibitemOpen
  \bibfield  {author} {\bibinfo {author} {\bibfnamefont {L.~P.}\ \bibnamefont
  {McGuinness}}, \bibinfo {author} {\bibfnamefont {Y.}~\bibnamefont {Yan}},
  \bibinfo {author} {\bibfnamefont {A.}~\bibnamefont {Stacey}}, \bibinfo
  {author} {\bibfnamefont {D.~A.}\ \bibnamefont {Simpson}}, \bibinfo {author}
  {\bibfnamefont {L.~T.}\ \bibnamefont {Hall}}, \bibinfo {author}
  {\bibfnamefont {D.}~\bibnamefont {Maclaurin}}, \bibinfo {author}
  {\bibfnamefont {S.}~\bibnamefont {Prawer}}, \bibinfo {author} {\bibfnamefont
  {P.}~\bibnamefont {Mulvaney}}, \bibinfo {author} {\bibfnamefont
  {J.}~\bibnamefont {Wrachtrup}}, \bibinfo {author} {\bibfnamefont
  {F.}~\bibnamefont {Caruso}}, \bibinfo {author} {\bibfnamefont {R.~E.}\
  \bibnamefont {Scholten}}, \ and\ \bibinfo {author} {\bibfnamefont {L.~C.~L.}\
  \bibnamefont {Hollenberg}},\ }\href {\doibase 10.1038/nnano.2011.64}
  {\bibfield  {journal} {\bibinfo  {journal} {Nature Nanotechnology}\ }\textbf
  {\bibinfo {volume} {6}},\ \bibinfo {pages} {358} (\bibinfo {year}
  {2011})}\BibitemShut {NoStop}%
\bibitem [{\citenamefont {McCoey}\ \emph {et~al.}(2020)\citenamefont {McCoey},
  \citenamefont {Matsuoka}, \citenamefont {Gille}, \citenamefont {Hall},
  \citenamefont {Shaw}, \citenamefont {Tetienne}, \citenamefont {Kisailus},
  \citenamefont {Hollenberg},\ and\ \citenamefont
  {Simpson}}]{mccoey_quantum_2020}%
  \BibitemOpen
  \bibfield  {author} {\bibinfo {author} {\bibfnamefont {J.~M.}\ \bibnamefont
  {McCoey}}, \bibinfo {author} {\bibfnamefont {M.}~\bibnamefont {Matsuoka}},
  \bibinfo {author} {\bibfnamefont {R.~W.}\ \bibnamefont {Gille}}, \bibinfo
  {author} {\bibfnamefont {L.~T.}\ \bibnamefont {Hall}}, \bibinfo {author}
  {\bibfnamefont {J.~A.}\ \bibnamefont {Shaw}}, \bibinfo {author}
  {\bibfnamefont {J.}~\bibnamefont {Tetienne}}, \bibinfo {author}
  {\bibfnamefont {D.}~\bibnamefont {Kisailus}}, \bibinfo {author}
  {\bibfnamefont {L.~C.~L.}\ \bibnamefont {Hollenberg}}, \ and\ \bibinfo
  {author} {\bibfnamefont {D.~A.}\ \bibnamefont {Simpson}},\ }\href {\doibase
  10.1002/smtd.201900754} {\bibfield  {journal} {\bibinfo  {journal} {Small
  Methods}\ }\textbf {\bibinfo {volume} {4}},\ \bibinfo {pages} {1900754}
  (\bibinfo {year} {2020})}\BibitemShut {NoStop}%
\bibitem [{\citenamefont {Ziem}\ \emph {et~al.}(2013)\citenamefont {Ziem},
  \citenamefont {Götz}, \citenamefont {Zappe}, \citenamefont {Steinert},\ and\
  \citenamefont {Wrachtrup}}]{ziem_highly_2013}%
  \BibitemOpen
  \bibfield  {author} {\bibinfo {author} {\bibfnamefont {F.~C.}\ \bibnamefont
  {Ziem}}, \bibinfo {author} {\bibfnamefont {N.~S.}\ \bibnamefont {Götz}},
  \bibinfo {author} {\bibfnamefont {A.}~\bibnamefont {Zappe}}, \bibinfo
  {author} {\bibfnamefont {S.}~\bibnamefont {Steinert}}, \ and\ \bibinfo
  {author} {\bibfnamefont {J.}~\bibnamefont {Wrachtrup}},\ }\href {\doibase
  10.1021/nl401522a} {\bibfield  {journal} {\bibinfo  {journal} {Nano Letters}\
  }\textbf {\bibinfo {volume} {13}},\ \bibinfo {pages} {4093} (\bibinfo {year}
  {2013})}\BibitemShut {NoStop}%
\bibitem [{\citenamefont {Ermakova}\ \emph {et~al.}(2013)\citenamefont
  {Ermakova}, \citenamefont {Pramanik}, \citenamefont {Cai}, \citenamefont
  {Algara-Siller}, \citenamefont {Kaiser}, \citenamefont {Weil}, \citenamefont
  {Tzeng}, \citenamefont {Chang}, \citenamefont {McGuinness}, \citenamefont
  {Plenio}, \citenamefont {Naydenov},\ and\ \citenamefont
  {Jelezko}}]{ermakova_detection_2013}%
  \BibitemOpen
  \bibfield  {author} {\bibinfo {author} {\bibfnamefont {A.}~\bibnamefont
  {Ermakova}}, \bibinfo {author} {\bibfnamefont {G.}~\bibnamefont {Pramanik}},
  \bibinfo {author} {\bibfnamefont {J.-M.}\ \bibnamefont {Cai}}, \bibinfo
  {author} {\bibfnamefont {G.}~\bibnamefont {Algara-Siller}}, \bibinfo {author}
  {\bibfnamefont {U.}~\bibnamefont {Kaiser}}, \bibinfo {author} {\bibfnamefont
  {T.}~\bibnamefont {Weil}}, \bibinfo {author} {\bibfnamefont {Y.-K.}\
  \bibnamefont {Tzeng}}, \bibinfo {author} {\bibfnamefont {H.~C.}\ \bibnamefont
  {Chang}}, \bibinfo {author} {\bibfnamefont {L.~P.}\ \bibnamefont
  {McGuinness}}, \bibinfo {author} {\bibfnamefont {M.~B.}\ \bibnamefont
  {Plenio}}, \bibinfo {author} {\bibfnamefont {B.}~\bibnamefont {Naydenov}}, \
  and\ \bibinfo {author} {\bibfnamefont {F.}~\bibnamefont {Jelezko}},\ }\href
  {\doibase 10.1021/nl4015233} {\bibfield  {journal} {\bibinfo  {journal} {Nano
  Letters}\ }\textbf {\bibinfo {volume} {13}},\ \bibinfo {pages} {3305}
  (\bibinfo {year} {2013})}\BibitemShut {NoStop}%
\bibitem [{\citenamefont {Barry}\ \emph {et~al.}(2016)\citenamefont {Barry},
  \citenamefont {Turner}, \citenamefont {Schloss}, \citenamefont {Glenn},
  \citenamefont {Song}, \citenamefont {Lukin}, \citenamefont {Park},\ and\
  \citenamefont {Walsworth}}]{barry_optical_2016}%
  \BibitemOpen
  \bibfield  {author} {\bibinfo {author} {\bibfnamefont {J.~F.}\ \bibnamefont
  {Barry}}, \bibinfo {author} {\bibfnamefont {M.~J.}\ \bibnamefont {Turner}},
  \bibinfo {author} {\bibfnamefont {J.~M.}\ \bibnamefont {Schloss}}, \bibinfo
  {author} {\bibfnamefont {D.~R.}\ \bibnamefont {Glenn}}, \bibinfo {author}
  {\bibfnamefont {Y.}~\bibnamefont {Song}}, \bibinfo {author} {\bibfnamefont
  {M.~D.}\ \bibnamefont {Lukin}}, \bibinfo {author} {\bibfnamefont
  {H.}~\bibnamefont {Park}}, \ and\ \bibinfo {author} {\bibfnamefont {R.~L.}\
  \bibnamefont {Walsworth}},\ }\href {\doibase 10.1073/pnas.1601513113}
  {\bibfield  {journal} {\bibinfo  {journal} {Proceedings of the National
  Academy of Sciences}\ }\textbf {\bibinfo {volume} {113}},\ \bibinfo {pages}
  {14133} (\bibinfo {year} {2016})}\BibitemShut {NoStop}%
\bibitem [{\citenamefont {DeVience}\ \emph {et~al.}(2015)\citenamefont
  {DeVience}, \citenamefont {Pham}, \citenamefont {Lovchinsky}, \citenamefont
  {Sushkov}, \citenamefont {Bar-Gill}, \citenamefont {Belthangady},
  \citenamefont {Casola}, \citenamefont {Corbett}, \citenamefont {Zhang},
  \citenamefont {Lukin}, \citenamefont {Park}, \citenamefont {Yacoby},\ and\
  \citenamefont {Walsworth}}]{devience_nanoscale_2015}%
  \BibitemOpen
  \bibfield  {author} {\bibinfo {author} {\bibfnamefont {S.~J.}\ \bibnamefont
  {DeVience}}, \bibinfo {author} {\bibfnamefont {L.~M.}\ \bibnamefont {Pham}},
  \bibinfo {author} {\bibfnamefont {I.}~\bibnamefont {Lovchinsky}}, \bibinfo
  {author} {\bibfnamefont {A.~O.}\ \bibnamefont {Sushkov}}, \bibinfo {author}
  {\bibfnamefont {N.}~\bibnamefont {Bar-Gill}}, \bibinfo {author}
  {\bibfnamefont {C.}~\bibnamefont {Belthangady}}, \bibinfo {author}
  {\bibfnamefont {F.}~\bibnamefont {Casola}}, \bibinfo {author} {\bibfnamefont
  {M.}~\bibnamefont {Corbett}}, \bibinfo {author} {\bibfnamefont
  {H.}~\bibnamefont {Zhang}}, \bibinfo {author} {\bibfnamefont
  {M.}~\bibnamefont {Lukin}}, \bibinfo {author} {\bibfnamefont
  {H.}~\bibnamefont {Park}}, \bibinfo {author} {\bibfnamefont {A.}~\bibnamefont
  {Yacoby}}, \ and\ \bibinfo {author} {\bibfnamefont {R.~L.}\ \bibnamefont
  {Walsworth}},\ }\href {\doibase 10.1038/nnano.2014.313} {\bibfield  {journal}
  {\bibinfo  {journal} {Nature Nanotechnology}\ }\textbf {\bibinfo {volume}
  {10}},\ \bibinfo {pages} {129} (\bibinfo {year} {2015})}\BibitemShut
  {NoStop}%
\bibitem [{\citenamefont {Rugar}\ \emph {et~al.}(2014)\citenamefont {Rugar},
  \citenamefont {Mamin}, \citenamefont {Sherwood}, \citenamefont {Kim},
  \citenamefont {Rettner}, \citenamefont {Ohno},\ and\ \citenamefont
  {Awschalom}}]{rugar_proton_2014}%
  \BibitemOpen
  \bibfield  {author} {\bibinfo {author} {\bibfnamefont {D.}~\bibnamefont
  {Rugar}}, \bibinfo {author} {\bibfnamefont {H.~J.}\ \bibnamefont {Mamin}},
  \bibinfo {author} {\bibfnamefont {M.~H.}\ \bibnamefont {Sherwood}}, \bibinfo
  {author} {\bibfnamefont {M.}~\bibnamefont {Kim}}, \bibinfo {author}
  {\bibfnamefont {C.~T.}\ \bibnamefont {Rettner}}, \bibinfo {author}
  {\bibfnamefont {K.}~\bibnamefont {Ohno}}, \ and\ \bibinfo {author}
  {\bibfnamefont {D.~D.}\ \bibnamefont {Awschalom}},\ }\href {\doibase
  10.1038/nnano.2014.288} {\bibfield  {journal} {\bibinfo  {journal} {Nature
  Nanotechnology}\ }\textbf {\bibinfo {volume} {10}},\ \bibinfo {pages} {120}
  (\bibinfo {year} {2014})}\BibitemShut {NoStop}%
\bibitem [{\citenamefont {Häberle}\ \emph {et~al.}(2015)\citenamefont
  {Häberle}, \citenamefont {Schmid-Lorch}, \citenamefont {Reinhard},\ and\
  \citenamefont {Wrachtrup}}]{haberle_nanoscale_2015}%
  \BibitemOpen
  \bibfield  {author} {\bibinfo {author} {\bibfnamefont {T.}~\bibnamefont
  {Häberle}}, \bibinfo {author} {\bibfnamefont {D.}~\bibnamefont
  {Schmid-Lorch}}, \bibinfo {author} {\bibfnamefont {F.}~\bibnamefont
  {Reinhard}}, \ and\ \bibinfo {author} {\bibfnamefont {J.}~\bibnamefont
  {Wrachtrup}},\ }\href {\doibase 10.1038/nnano.2014.299} {\bibfield  {journal}
  {\bibinfo  {journal} {Nature Nanotechnology}\ }\textbf {\bibinfo {volume}
  {10}},\ \bibinfo {pages} {125} (\bibinfo {year} {2015})}\BibitemShut
  {NoStop}%
\bibitem [{\citenamefont {Mamin}\ \emph {et~al.}(2013)\citenamefont {Mamin},
  \citenamefont {Kim}, \citenamefont {Sherwood}, \citenamefont {Rettner},
  \citenamefont {Ohno}, \citenamefont {Awschalom},\ and\ \citenamefont
  {Rugar}}]{mamin_nanoscale_2013}%
  \BibitemOpen
  \bibfield  {author} {\bibinfo {author} {\bibfnamefont {H.~J.}\ \bibnamefont
  {Mamin}}, \bibinfo {author} {\bibfnamefont {M.}~\bibnamefont {Kim}}, \bibinfo
  {author} {\bibfnamefont {M.~H.}\ \bibnamefont {Sherwood}}, \bibinfo {author}
  {\bibfnamefont {C.~T.}\ \bibnamefont {Rettner}}, \bibinfo {author}
  {\bibfnamefont {K.}~\bibnamefont {Ohno}}, \bibinfo {author} {\bibfnamefont
  {D.~D.}\ \bibnamefont {Awschalom}}, \ and\ \bibinfo {author} {\bibfnamefont
  {D.}~\bibnamefont {Rugar}},\ }\href {\doibase 10.1126/science.1231540}
  {\bibfield  {journal} {\bibinfo  {journal} {Science}\ }\textbf {\bibinfo
  {volume} {339}},\ \bibinfo {pages} {557} (\bibinfo {year}
  {2013})}\BibitemShut {NoStop}%
\bibitem [{\citenamefont {Bhallamudi}\ and\ \citenamefont
  {Hammel}(2015)}]{bhallamudi_nitrogen-vacancy_2015}%
  \BibitemOpen
  \bibfield  {author} {\bibinfo {author} {\bibfnamefont {V.~P.}\ \bibnamefont
  {Bhallamudi}}\ and\ \bibinfo {author} {\bibfnamefont {P.~C.}\ \bibnamefont
  {Hammel}},\ }\href
  {http://www.nature.com/nnano/journal/v10/n2/full/nnano.2015.7.html}
  {\bibfield  {journal} {\bibinfo  {journal} {Nature nanotechnology}\ }\textbf
  {\bibinfo {volume} {10}},\ \bibinfo {pages} {104} (\bibinfo {year}
  {2015})}\BibitemShut {NoStop}%
\bibitem [{\citenamefont {Staudacher}\ \emph {et~al.}(2013)\citenamefont
  {Staudacher}, \citenamefont {Shi}, \citenamefont {Pezzagna}, \citenamefont
  {Meijer}, \citenamefont {Du}, \citenamefont {Meriles}, \citenamefont
  {Reinhard},\ and\ \citenamefont {Wrachtrup}}]{staudacher_nuclear_2013}%
  \BibitemOpen
  \bibfield  {author} {\bibinfo {author} {\bibfnamefont {T.}~\bibnamefont
  {Staudacher}}, \bibinfo {author} {\bibfnamefont {F.}~\bibnamefont {Shi}},
  \bibinfo {author} {\bibfnamefont {S.}~\bibnamefont {Pezzagna}}, \bibinfo
  {author} {\bibfnamefont {J.}~\bibnamefont {Meijer}}, \bibinfo {author}
  {\bibfnamefont {J.}~\bibnamefont {Du}}, \bibinfo {author} {\bibfnamefont
  {C.~A.}\ \bibnamefont {Meriles}}, \bibinfo {author} {\bibfnamefont
  {F.}~\bibnamefont {Reinhard}}, \ and\ \bibinfo {author} {\bibfnamefont
  {J.}~\bibnamefont {Wrachtrup}},\ }\href {\doibase 10.1126/science.1231675}
  {\bibfield  {journal} {\bibinfo  {journal} {Science}\ }\textbf {\bibinfo
  {volume} {339}},\ \bibinfo {pages} {561} (\bibinfo {year}
  {2013})}\BibitemShut {NoStop}%
\bibitem [{\citenamefont {Jahnke}\ \emph {et~al.}(2012)\citenamefont {Jahnke},
  \citenamefont {Naydenov}, \citenamefont {Teraji}, \citenamefont {Koizumi},
  \citenamefont {Umeda}, \citenamefont {Isoya},\ and\ \citenamefont
  {Jelezko}}]{jahnke_long_2012}%
  \BibitemOpen
  \bibfield  {author} {\bibinfo {author} {\bibfnamefont {K.~D.}\ \bibnamefont
  {Jahnke}}, \bibinfo {author} {\bibfnamefont {B.}~\bibnamefont {Naydenov}},
  \bibinfo {author} {\bibfnamefont {T.}~\bibnamefont {Teraji}}, \bibinfo
  {author} {\bibfnamefont {S.}~\bibnamefont {Koizumi}}, \bibinfo {author}
  {\bibfnamefont {T.}~\bibnamefont {Umeda}}, \bibinfo {author} {\bibfnamefont
  {J.}~\bibnamefont {Isoya}}, \ and\ \bibinfo {author} {\bibfnamefont
  {F.}~\bibnamefont {Jelezko}},\ }\href {\doibase 10.1063/1.4731778} {\bibfield
   {journal} {\bibinfo  {journal} {Applied Physics Letters}\ }\textbf {\bibinfo
  {volume} {101}},\ \bibinfo {pages} {012405} (\bibinfo {year}
  {2012})}\BibitemShut {NoStop}%
\bibitem [{\citenamefont {Balasubramanian}\ \emph {et~al.}(2009)\citenamefont
  {Balasubramanian}, \citenamefont {Neumann}, \citenamefont {Twitchen},
  \citenamefont {Markham}, \citenamefont {Kolesov}, \citenamefont {Mizuochi},
  \citenamefont {Isoya}, \citenamefont {Achard}, \citenamefont {Beck},
  \citenamefont {Tissler}, \citenamefont {Jacques}, \citenamefont {Hemmer},
  \citenamefont {Jelezko},\ and\ \citenamefont
  {Wrachtrup}}]{balasubramanian_ultralong_2009}%
  \BibitemOpen
  \bibfield  {author} {\bibinfo {author} {\bibfnamefont {G.}~\bibnamefont
  {Balasubramanian}}, \bibinfo {author} {\bibfnamefont {P.}~\bibnamefont
  {Neumann}}, \bibinfo {author} {\bibfnamefont {D.}~\bibnamefont {Twitchen}},
  \bibinfo {author} {\bibfnamefont {M.}~\bibnamefont {Markham}}, \bibinfo
  {author} {\bibfnamefont {R.}~\bibnamefont {Kolesov}}, \bibinfo {author}
  {\bibfnamefont {N.}~\bibnamefont {Mizuochi}}, \bibinfo {author}
  {\bibfnamefont {J.}~\bibnamefont {Isoya}}, \bibinfo {author} {\bibfnamefont
  {J.}~\bibnamefont {Achard}}, \bibinfo {author} {\bibfnamefont
  {J.}~\bibnamefont {Beck}}, \bibinfo {author} {\bibfnamefont {J.}~\bibnamefont
  {Tissler}}, \bibinfo {author} {\bibfnamefont {V.}~\bibnamefont {Jacques}},
  \bibinfo {author} {\bibfnamefont {P.~R.}\ \bibnamefont {Hemmer}}, \bibinfo
  {author} {\bibfnamefont {F.}~\bibnamefont {Jelezko}}, \ and\ \bibinfo
  {author} {\bibfnamefont {J.}~\bibnamefont {Wrachtrup}},\ }\href {\doibase
  10.1038/nmat2420} {\bibfield  {journal} {\bibinfo  {journal} {Nature
  Materials}\ }\textbf {\bibinfo {volume} {8}},\ \bibinfo {pages} {383}
  (\bibinfo {year} {2009})}\BibitemShut {NoStop}%
\bibitem [{\citenamefont {Hadden}\ \emph {et~al.}(2010)\citenamefont {Hadden},
  \citenamefont {Harrison}, \citenamefont {Stanley-Clarke}, \citenamefont
  {Marseglia}, \citenamefont {Ho}, \citenamefont {Patton}, \citenamefont
  {O’Brien},\ and\ \citenamefont {Rarity}}]{hadden_strongly_2010}%
  \BibitemOpen
  \bibfield  {author} {\bibinfo {author} {\bibfnamefont {J.~P.}\ \bibnamefont
  {Hadden}}, \bibinfo {author} {\bibfnamefont {J.~P.}\ \bibnamefont
  {Harrison}}, \bibinfo {author} {\bibfnamefont {A.~C.}\ \bibnamefont
  {Stanley-Clarke}}, \bibinfo {author} {\bibfnamefont {L.}~\bibnamefont
  {Marseglia}}, \bibinfo {author} {\bibfnamefont {Y.-L.~D.}\ \bibnamefont
  {Ho}}, \bibinfo {author} {\bibfnamefont {B.~R.}\ \bibnamefont {Patton}},
  \bibinfo {author} {\bibfnamefont {J.~L.}\ \bibnamefont {O’Brien}}, \ and\
  \bibinfo {author} {\bibfnamefont {J.~G.}\ \bibnamefont {Rarity}},\ }\href
  {\doibase 10.1063/1.3519847} {\bibfield  {journal} {\bibinfo  {journal}
  {Applied Physics Letters}\ }\textbf {\bibinfo {volume} {97}},\ \bibinfo
  {pages} {241901} (\bibinfo {year} {2010})}\BibitemShut {NoStop}%
\bibitem [{\citenamefont {Jensen}\ \emph {et~al.}(2014)\citenamefont {Jensen},
  \citenamefont {Leefer}, \citenamefont {Jarmola}, \citenamefont {Dumeige},
  \citenamefont {Acosta}, \citenamefont {Kehayias}, \citenamefont {Patton},\
  and\ \citenamefont {Budker}}]{jensen_cavity-enhanced_2014}%
  \BibitemOpen
  \bibfield  {author} {\bibinfo {author} {\bibfnamefont {K.}~\bibnamefont
  {Jensen}}, \bibinfo {author} {\bibfnamefont {N.}~\bibnamefont {Leefer}},
  \bibinfo {author} {\bibfnamefont {A.}~\bibnamefont {Jarmola}}, \bibinfo
  {author} {\bibfnamefont {Y.}~\bibnamefont {Dumeige}}, \bibinfo {author}
  {\bibfnamefont {V.}~\bibnamefont {Acosta}}, \bibinfo {author} {\bibfnamefont
  {P.}~\bibnamefont {Kehayias}}, \bibinfo {author} {\bibfnamefont
  {B.}~\bibnamefont {Patton}}, \ and\ \bibinfo {author} {\bibfnamefont
  {D.}~\bibnamefont {Budker}},\ }\href {\doibase
  10.1103/PhysRevLett.112.160802} {\bibfield  {journal} {\bibinfo  {journal}
  {Physical Review Letters}\ }\textbf {\bibinfo {volume} {112}},\ \bibinfo
  {pages} {160802} (\bibinfo {year} {2014})}\BibitemShut {NoStop}%
\bibitem [{\citenamefont {Riedrich-Möller}\ \emph {et~al.}(2015)\citenamefont
  {Riedrich-Möller}, \citenamefont {Pezzagna}, \citenamefont {Meijer},
  \citenamefont {Pauly}, \citenamefont {Mücklich}, \citenamefont {Markham},
  \citenamefont {Edmonds},\ and\ \citenamefont
  {Becher}}]{riedrich-moller_nanoimplantation_2015}%
  \BibitemOpen
  \bibfield  {author} {\bibinfo {author} {\bibfnamefont {J.}~\bibnamefont
  {Riedrich-Möller}}, \bibinfo {author} {\bibfnamefont {S.}~\bibnamefont
  {Pezzagna}}, \bibinfo {author} {\bibfnamefont {J.}~\bibnamefont {Meijer}},
  \bibinfo {author} {\bibfnamefont {C.}~\bibnamefont {Pauly}}, \bibinfo
  {author} {\bibfnamefont {F.}~\bibnamefont {Mücklich}}, \bibinfo {author}
  {\bibfnamefont {M.}~\bibnamefont {Markham}}, \bibinfo {author} {\bibfnamefont
  {A.~M.}\ \bibnamefont {Edmonds}}, \ and\ \bibinfo {author} {\bibfnamefont
  {C.}~\bibnamefont {Becher}},\ }\href {\doibase 10.1063/1.4922117} {\bibfield
  {journal} {\bibinfo  {journal} {Applied Physics Letters}\ }\textbf {\bibinfo
  {volume} {106}},\ \bibinfo {pages} {221103} (\bibinfo {year}
  {2015})}\BibitemShut {NoStop}%
\bibitem [{\citenamefont {Babinec}\ \emph {et~al.}(2010)\citenamefont
  {Babinec}, \citenamefont {Hausmann}, \citenamefont {Khan}, \citenamefont
  {Zhang}, \citenamefont {Maze}, \citenamefont {Hemmer},\ and\ \citenamefont
  {Lončar}}]{babinec_diamond_2010}%
  \BibitemOpen
  \bibfield  {author} {\bibinfo {author} {\bibfnamefont {T.~M.}\ \bibnamefont
  {Babinec}}, \bibinfo {author} {\bibfnamefont {B.~J.~M.}\ \bibnamefont
  {Hausmann}}, \bibinfo {author} {\bibfnamefont {M.}~\bibnamefont {Khan}},
  \bibinfo {author} {\bibfnamefont {Y.}~\bibnamefont {Zhang}}, \bibinfo
  {author} {\bibfnamefont {J.~R.}\ \bibnamefont {Maze}}, \bibinfo {author}
  {\bibfnamefont {P.~R.}\ \bibnamefont {Hemmer}}, \ and\ \bibinfo {author}
  {\bibfnamefont {M.}~\bibnamefont {Lončar}},\ }\href {\doibase
  10.1038/nnano.2010.6} {\bibfield  {journal} {\bibinfo  {journal} {Nature
  Nanotechnology}\ }\textbf {\bibinfo {volume} {5}},\ \bibinfo {pages} {195}
  (\bibinfo {year} {2010})}\BibitemShut {NoStop}%
\bibitem [{\citenamefont {Maletinsky}\ \emph {et~al.}(2012)\citenamefont
  {Maletinsky}, \citenamefont {Hong}, \citenamefont {Grinolds}, \citenamefont
  {Hausmann}, \citenamefont {Lukin}, \citenamefont {Walsworth}, \citenamefont
  {Loncar},\ and\ \citenamefont {Yacoby}}]{maletinsky_robust_2012}%
  \BibitemOpen
  \bibfield  {author} {\bibinfo {author} {\bibfnamefont {P.}~\bibnamefont
  {Maletinsky}}, \bibinfo {author} {\bibfnamefont {S.}~\bibnamefont {Hong}},
  \bibinfo {author} {\bibfnamefont {M.~S.}\ \bibnamefont {Grinolds}}, \bibinfo
  {author} {\bibfnamefont {B.}~\bibnamefont {Hausmann}}, \bibinfo {author}
  {\bibfnamefont {M.~D.}\ \bibnamefont {Lukin}}, \bibinfo {author}
  {\bibfnamefont {R.~L.}\ \bibnamefont {Walsworth}}, \bibinfo {author}
  {\bibfnamefont {M.}~\bibnamefont {Loncar}}, \ and\ \bibinfo {author}
  {\bibfnamefont {A.}~\bibnamefont {Yacoby}},\ }\href {\doibase
  10.1038/nnano.2012.50} {\bibfield  {journal} {\bibinfo  {journal} {Nature
  Nanotechnology}\ }\textbf {\bibinfo {volume} {7}},\ \bibinfo {pages} {320}
  (\bibinfo {year} {2012})}\BibitemShut {NoStop}%
\bibitem [{\citenamefont {Pelliccione}\ \emph {et~al.}(2016)\citenamefont
  {Pelliccione}, \citenamefont {Jenkins}, \citenamefont {Ovartchaiyapong},
  \citenamefont {Reetz}, \citenamefont {Emmanouilidou}, \citenamefont {Ni},\
  and\ \citenamefont {Bleszynski~Jayich}}]{pelliccione_scanned_2016}%
  \BibitemOpen
  \bibfield  {author} {\bibinfo {author} {\bibfnamefont {M.}~\bibnamefont
  {Pelliccione}}, \bibinfo {author} {\bibfnamefont {A.}~\bibnamefont
  {Jenkins}}, \bibinfo {author} {\bibfnamefont {P.}~\bibnamefont
  {Ovartchaiyapong}}, \bibinfo {author} {\bibfnamefont {C.}~\bibnamefont
  {Reetz}}, \bibinfo {author} {\bibfnamefont {E.}~\bibnamefont
  {Emmanouilidou}}, \bibinfo {author} {\bibfnamefont {N.}~\bibnamefont {Ni}}, \
  and\ \bibinfo {author} {\bibfnamefont {A.~C.}\ \bibnamefont
  {Bleszynski~Jayich}},\ }\href {\doibase 10.1038/nnano.2016.68} {\bibfield
  {journal} {\bibinfo  {journal} {Nature Nanotechnology}\ }\textbf {\bibinfo
  {volume} {11}},\ \bibinfo {pages} {700} (\bibinfo {year} {2016})}\BibitemShut
  {NoStop}%
\bibitem [{\citenamefont {Dinani}\ \emph {et~al.}(2019)\citenamefont {Dinani},
  \citenamefont {Berry}, \citenamefont {Gonzalez}, \citenamefont {Maze},\ and\
  \citenamefont {Bonato}}]{dinani_bayesian_2019}%
  \BibitemOpen
  \bibfield  {author} {\bibinfo {author} {\bibfnamefont {H.~T.}\ \bibnamefont
  {Dinani}}, \bibinfo {author} {\bibfnamefont {D.~W.}\ \bibnamefont {Berry}},
  \bibinfo {author} {\bibfnamefont {R.}~\bibnamefont {Gonzalez}}, \bibinfo
  {author} {\bibfnamefont {J.~R.}\ \bibnamefont {Maze}}, \ and\ \bibinfo
  {author} {\bibfnamefont {C.}~\bibnamefont {Bonato}},\ }\href {\doibase
  10.1103/PhysRevB.99.125413} {\bibfield  {journal} {\bibinfo  {journal}
  {Physical Review B}\ }\textbf {\bibinfo {volume} {99}},\ \bibinfo {pages}
  {125413} (\bibinfo {year} {2019})}\BibitemShut {NoStop}%
\bibitem [{\citenamefont {Waldherr}\ \emph {et~al.}(2011)\citenamefont
  {Waldherr}, \citenamefont {Neumann}, \citenamefont {Huelga}, \citenamefont
  {Jelezko},\ and\ \citenamefont {Wrachtrup}}]{waldherr_violation_2011}%
  \BibitemOpen
  \bibfield  {author} {\bibinfo {author} {\bibfnamefont {G.}~\bibnamefont
  {Waldherr}}, \bibinfo {author} {\bibfnamefont {P.}~\bibnamefont {Neumann}},
  \bibinfo {author} {\bibfnamefont {S.~F.}\ \bibnamefont {Huelga}}, \bibinfo
  {author} {\bibfnamefont {F.}~\bibnamefont {Jelezko}}, \ and\ \bibinfo
  {author} {\bibfnamefont {J.}~\bibnamefont {Wrachtrup}},\ }\href {\doibase
  10.1103/PhysRevLett.107.090401} {\bibfield  {journal} {\bibinfo  {journal}
  {Physical Review Letters}\ }\textbf {\bibinfo {volume} {107}},\ \bibinfo
  {pages} {090401} (\bibinfo {year} {2011})}\BibitemShut {NoStop}%
\bibitem [{\citenamefont {Aslam}\ \emph {et~al.}(2013)\citenamefont {Aslam},
  \citenamefont {Waldherr}, \citenamefont {Neumann}, \citenamefont {Jelezko},\
  and\ \citenamefont {Wrachtrup}}]{aslam_photo-induced_2013}%
  \BibitemOpen
  \bibfield  {author} {\bibinfo {author} {\bibfnamefont {N.}~\bibnamefont
  {Aslam}}, \bibinfo {author} {\bibfnamefont {G.}~\bibnamefont {Waldherr}},
  \bibinfo {author} {\bibfnamefont {P.}~\bibnamefont {Neumann}}, \bibinfo
  {author} {\bibfnamefont {F.}~\bibnamefont {Jelezko}}, \ and\ \bibinfo
  {author} {\bibfnamefont {J.}~\bibnamefont {Wrachtrup}},\ }\href {\doibase
  10.1088/1367-2630/15/1/013064} {\bibfield  {journal} {\bibinfo  {journal}
  {New Journal of Physics}\ }\textbf {\bibinfo {volume} {15}},\ \bibinfo
  {pages} {013064} (\bibinfo {year} {2013})}\BibitemShut {NoStop}%
\bibitem [{\citenamefont {Shields}\ \emph {et~al.}(2015)\citenamefont
  {Shields}, \citenamefont {Unterreithmeier}, \citenamefont {de~Leon},
  \citenamefont {Park},\ and\ \citenamefont {Lukin}}]{shields_efficient_2015}%
  \BibitemOpen
  \bibfield  {author} {\bibinfo {author} {\bibfnamefont {B.}~\bibnamefont
  {Shields}}, \bibinfo {author} {\bibfnamefont {Q.}~\bibnamefont
  {Unterreithmeier}}, \bibinfo {author} {\bibfnamefont {N.}~\bibnamefont
  {de~Leon}}, \bibinfo {author} {\bibfnamefont {H.}~\bibnamefont {Park}}, \
  and\ \bibinfo {author} {\bibfnamefont {M.}~\bibnamefont {Lukin}},\ }\href
  {\doibase 10.1103/PhysRevLett.114.136402} {\bibfield  {journal} {\bibinfo
  {journal} {Physical Review Letters}\ }\textbf {\bibinfo {volume} {114}},\
  \bibinfo {pages} {136402} (\bibinfo {year} {2015})}\BibitemShut {NoStop}%
\bibitem [{\citenamefont {Bluvstein}, \citenamefont {Zhang},\ and\
  \citenamefont {Jayich}(2019)}]{bluvstein_identifying_2019}%
  \BibitemOpen
  \bibfield  {author} {\bibinfo {author} {\bibfnamefont {D.}~\bibnamefont
  {Bluvstein}}, \bibinfo {author} {\bibfnamefont {Z.}~\bibnamefont {Zhang}}, \
  and\ \bibinfo {author} {\bibfnamefont {A.~C.~B.}\ \bibnamefont {Jayich}},\
  }\href@noop {} {\bibfield  {journal} {\bibinfo  {journal} {Physical Review
  Letters}\ }\textbf {\bibinfo {volume} {122}},\ \bibinfo {pages} {076101}
  (\bibinfo {year} {2019})}\BibitemShut {NoStop}%
\bibitem [{\citenamefont {Jaskula}\ \emph {et~al.}(2019)\citenamefont
  {Jaskula}, \citenamefont {Shields}, \citenamefont {Bauch}, \citenamefont
  {Lukin}, \citenamefont {Trifonov},\ and\ \citenamefont
  {Walsworth}}]{jaskula_improved_2019}%
  \BibitemOpen
  \bibfield  {author} {\bibinfo {author} {\bibfnamefont {J.-C.}\ \bibnamefont
  {Jaskula}}, \bibinfo {author} {\bibfnamefont {B.}~\bibnamefont {Shields}},
  \bibinfo {author} {\bibfnamefont {E.}~\bibnamefont {Bauch}}, \bibinfo
  {author} {\bibfnamefont {M.}~\bibnamefont {Lukin}}, \bibinfo {author}
  {\bibfnamefont {A.}~\bibnamefont {Trifonov}}, \ and\ \bibinfo {author}
  {\bibfnamefont {R.}~\bibnamefont {Walsworth}},\ }\href {\doibase
  10.1103/PhysRevApplied.11.064003} {\bibfield  {journal} {\bibinfo  {journal}
  {Physical Review Applied}\ }\textbf {\bibinfo {volume} {11}},\ \bibinfo
  {pages} {064003} (\bibinfo {year} {2019})}\BibitemShut {NoStop}%
\bibitem [{\citenamefont {Hacquebard}\ and\ \citenamefont
  {Childress}(2018)}]{hacquebard_charge-state_2018}%
  \BibitemOpen
  \bibfield  {author} {\bibinfo {author} {\bibfnamefont {L.}~\bibnamefont
  {Hacquebard}}\ and\ \bibinfo {author} {\bibfnamefont {L.}~\bibnamefont
  {Childress}},\ }\href {\doibase 10.1103/PhysRevA.97.063408} {\bibfield
  {journal} {\bibinfo  {journal} {Physical Review A}\ }\textbf {\bibinfo
  {volume} {97}},\ \bibinfo {pages} {063408} (\bibinfo {year}
  {2018})}\BibitemShut {NoStop}%
\bibitem [{\citenamefont {Hopper}\ \emph {et~al.}(2016)\citenamefont {Hopper},
  \citenamefont {Grote}, \citenamefont {Exarhos},\ and\ \citenamefont
  {Bassett}}]{hopper_near-infrared-assisted_2016}%
  \BibitemOpen
  \bibfield  {author} {\bibinfo {author} {\bibfnamefont {D.~A.}\ \bibnamefont
  {Hopper}}, \bibinfo {author} {\bibfnamefont {R.~R.}\ \bibnamefont {Grote}},
  \bibinfo {author} {\bibfnamefont {A.~L.}\ \bibnamefont {Exarhos}}, \ and\
  \bibinfo {author} {\bibfnamefont {L.~C.}\ \bibnamefont {Bassett}},\ }\href
  {\doibase 10.1103/PhysRevB.94.241201} {\bibfield  {journal} {\bibinfo
  {journal} {Physical Review B}\ }\textbf {\bibinfo {volume} {94}},\ \bibinfo
  {pages} {241201} (\bibinfo {year} {2016})}\BibitemShut {NoStop}%
\bibitem [{\citenamefont {Acosta}\ \emph
  {et~al.}(2010{\natexlab{a}})\citenamefont {Acosta}, \citenamefont {Jarmola},
  \citenamefont {Bauch},\ and\ \citenamefont {Budker}}]{acosta_optical_2010}%
  \BibitemOpen
  \bibfield  {author} {\bibinfo {author} {\bibfnamefont {V.~M.}\ \bibnamefont
  {Acosta}}, \bibinfo {author} {\bibfnamefont {A.}~\bibnamefont {Jarmola}},
  \bibinfo {author} {\bibfnamefont {E.}~\bibnamefont {Bauch}}, \ and\ \bibinfo
  {author} {\bibfnamefont {D.}~\bibnamefont {Budker}},\ }\href {\doibase
  10.1103/PhysRevB.82.201202} {\bibfield  {journal} {\bibinfo  {journal}
  {Physical Review B}\ }\textbf {\bibinfo {volume} {82}},\ \bibinfo {pages}
  {201202} (\bibinfo {year} {2010}{\natexlab{a}})}\BibitemShut {NoStop}%
\bibitem [{\citenamefont {Acosta}\ \emph
  {et~al.}(2010{\natexlab{b}})\citenamefont {Acosta}, \citenamefont {Bauch},
  \citenamefont {Jarmola}, \citenamefont {Zipp}, \citenamefont {Ledbetter},\
  and\ \citenamefont {Budker}}]{acosta_broadband_2010}%
  \BibitemOpen
  \bibfield  {author} {\bibinfo {author} {\bibfnamefont {V.~M.}\ \bibnamefont
  {Acosta}}, \bibinfo {author} {\bibfnamefont {E.}~\bibnamefont {Bauch}},
  \bibinfo {author} {\bibfnamefont {A.}~\bibnamefont {Jarmola}}, \bibinfo
  {author} {\bibfnamefont {L.~J.}\ \bibnamefont {Zipp}}, \bibinfo {author}
  {\bibfnamefont {M.~P.}\ \bibnamefont {Ledbetter}}, \ and\ \bibinfo {author}
  {\bibfnamefont {D.}~\bibnamefont {Budker}},\ }\href {\doibase
  10.1063/1.3507884} {\bibfield  {journal} {\bibinfo  {journal} {Applied
  Physics Letters}\ }\textbf {\bibinfo {volume} {97}},\ \bibinfo {pages}
  {174104} (\bibinfo {year} {2010}{\natexlab{b}})}\BibitemShut {NoStop}%
\bibitem [{\citenamefont {Higgins}\ \emph {et~al.}(2007)\citenamefont
  {Higgins}, \citenamefont {Berry}, \citenamefont {Bartlett}, \citenamefont
  {Wiseman},\ and\ \citenamefont {Pryde}}]{higgins_entanglement-free_2007}%
  \BibitemOpen
  \bibfield  {author} {\bibinfo {author} {\bibfnamefont {B.~L.}\ \bibnamefont
  {Higgins}}, \bibinfo {author} {\bibfnamefont {D.~W.}\ \bibnamefont {Berry}},
  \bibinfo {author} {\bibfnamefont {S.~D.}\ \bibnamefont {Bartlett}}, \bibinfo
  {author} {\bibfnamefont {H.~M.}\ \bibnamefont {Wiseman}}, \ and\ \bibinfo
  {author} {\bibfnamefont {G.~J.}\ \bibnamefont {Pryde}},\ }\href {\doibase
  10.1038/nature06257} {\bibfield  {journal} {\bibinfo  {journal} {Nature}\
  }\textbf {\bibinfo {volume} {450}},\ \bibinfo {pages} {393} (\bibinfo {year}
  {2007})}\BibitemShut {NoStop}%
\bibitem [{\citenamefont {Berry}\ \emph {et~al.}(2009)\citenamefont {Berry},
  \citenamefont {Higgins}, \citenamefont {Bartlett}, \citenamefont {Mitchell},
  \citenamefont {Pryde},\ and\ \citenamefont {Wiseman}}]{berry_how_2009}%
  \BibitemOpen
  \bibfield  {author} {\bibinfo {author} {\bibfnamefont {D.~W.}\ \bibnamefont
  {Berry}}, \bibinfo {author} {\bibfnamefont {B.~L.}\ \bibnamefont {Higgins}},
  \bibinfo {author} {\bibfnamefont {S.~D.}\ \bibnamefont {Bartlett}}, \bibinfo
  {author} {\bibfnamefont {M.~W.}\ \bibnamefont {Mitchell}}, \bibinfo {author}
  {\bibfnamefont {G.~J.}\ \bibnamefont {Pryde}}, \ and\ \bibinfo {author}
  {\bibfnamefont {H.~M.}\ \bibnamefont {Wiseman}},\ }\href {\doibase
  10.1103/PhysRevA.80.052114} {\bibfield  {journal} {\bibinfo  {journal}
  {Physical Review A}\ }\textbf {\bibinfo {volume} {80}},\ \bibinfo {pages}
  {052114} (\bibinfo {year} {2009})}\BibitemShut {NoStop}%
\bibitem [{\citenamefont {Said}, \citenamefont {Berry},\ and\ \citenamefont
  {Twamley}(2011)}]{said_nanoscale_2011}%
  \BibitemOpen
  \bibfield  {author} {\bibinfo {author} {\bibfnamefont {R.~S.}\ \bibnamefont
  {Said}}, \bibinfo {author} {\bibfnamefont {D.~W.}\ \bibnamefont {Berry}}, \
  and\ \bibinfo {author} {\bibfnamefont {J.}~\bibnamefont {Twamley}},\
  }\href@noop {} {\bibfield  {journal} {\bibinfo  {journal} {Physical Review
  B}\ }\textbf {\bibinfo {volume} {83}},\ \bibinfo {pages} {125410} (\bibinfo
  {year} {2011})}\BibitemShut {NoStop}%
\bibitem [{\citenamefont {Cappellaro}(2012)}]{cappellaro_spin-bath_2012}%
  \BibitemOpen
  \bibfield  {author} {\bibinfo {author} {\bibfnamefont {P.}~\bibnamefont
  {Cappellaro}},\ }\href {\doibase 10.1103/PhysRevA.85.030301} {\bibfield
  {journal} {\bibinfo  {journal} {Physical Review A}\ }\textbf {\bibinfo
  {volume} {85}},\ \bibinfo {pages} {030301} (\bibinfo {year}
  {2012})}\BibitemShut {NoStop}%
\bibitem [{\citenamefont {Degen}, \citenamefont {Reinhard},\ and\ \citenamefont
  {Cappellaro}(2017)}]{degen_quantum_2017}%
  \BibitemOpen
  \bibfield  {author} {\bibinfo {author} {\bibfnamefont {C.}~\bibnamefont
  {Degen}}, \bibinfo {author} {\bibfnamefont {F.}~\bibnamefont {Reinhard}}, \
  and\ \bibinfo {author} {\bibfnamefont {P.}~\bibnamefont {Cappellaro}},\
  }\href {\doibase 10.1103/RevModPhys.89.035002} {\bibfield  {journal}
  {\bibinfo  {journal} {Reviews of Modern Physics}\ }\textbf {\bibinfo {volume}
  {89}},\ \bibinfo {pages} {035002} (\bibinfo {year} {2017})}\BibitemShut
  {NoStop}%
\bibitem [{\citenamefont {Gentile}\ \emph {et~al.}(2019)\citenamefont
  {Gentile}, \citenamefont {Santagati}, \citenamefont {Schmitt},\ and\
  \citenamefont {McGuinness}}]{gentile_experimental_2019}%
  \BibitemOpen
  \bibfield  {author} {\bibinfo {author} {\bibfnamefont {A.}~\bibnamefont
  {Gentile}}, \bibinfo {author} {\bibfnamefont {R.}~\bibnamefont {Santagati}},
  \bibinfo {author} {\bibfnamefont {S.}~\bibnamefont {Schmitt}}, \ and\
  \bibinfo {author} {\bibfnamefont {L.~P.}\ \bibnamefont {McGuinness}},\ }\href
  {\doibase 10.6084/M9.FIGSHARE.7855118} {\enquote {\bibinfo {title}
  {Experimental {Datasets}},}\ } (\bibinfo {year} {2019})\BibitemShut {NoStop}%
\bibitem [{\citenamefont {Santagati}\ \emph {et~al.}(2019)\citenamefont
  {Santagati}, \citenamefont {Gentile}, \citenamefont {Knauer}, \citenamefont
  {Schmitt}, \citenamefont {Paesani}, \citenamefont {Granade}, \citenamefont
  {Wiebe}, \citenamefont {Osterkamp}, \citenamefont {McGuinness}, \citenamefont
  {Wang}, \citenamefont {Thompson}, \citenamefont {Rarity}, \citenamefont
  {Jelezko},\ and\ \citenamefont {Laing}}]{santagati_magnetic-field_2019}%
  \BibitemOpen
  \bibfield  {author} {\bibinfo {author} {\bibfnamefont {R.}~\bibnamefont
  {Santagati}}, \bibinfo {author} {\bibfnamefont {A.}~\bibnamefont {Gentile}},
  \bibinfo {author} {\bibfnamefont {S.}~\bibnamefont {Knauer}}, \bibinfo
  {author} {\bibfnamefont {S.}~\bibnamefont {Schmitt}}, \bibinfo {author}
  {\bibfnamefont {S.}~\bibnamefont {Paesani}}, \bibinfo {author} {\bibfnamefont
  {C.}~\bibnamefont {Granade}}, \bibinfo {author} {\bibfnamefont
  {N.}~\bibnamefont {Wiebe}}, \bibinfo {author} {\bibfnamefont
  {C.}~\bibnamefont {Osterkamp}}, \bibinfo {author} {\bibfnamefont
  {L.}~\bibnamefont {McGuinness}}, \bibinfo {author} {\bibfnamefont
  {J.}~\bibnamefont {Wang}}, \bibinfo {author} {\bibfnamefont {M.}~\bibnamefont
  {Thompson}}, \bibinfo {author} {\bibfnamefont {J.}~\bibnamefont {Rarity}},
  \bibinfo {author} {\bibfnamefont {F.}~\bibnamefont {Jelezko}}, \ and\
  \bibinfo {author} {\bibfnamefont {A.}~\bibnamefont {Laing}},\ }\href
  {\doibase 10.1103/PhysRevX.9.021019} {\bibfield  {journal} {\bibinfo
  {journal} {Physical Review X}\ }\textbf {\bibinfo {volume} {9}},\ \bibinfo
  {pages} {021019} (\bibinfo {year} {2019})}\BibitemShut {NoStop}%
\bibitem [{\citenamefont {Chaloner}\ and\ \citenamefont
  {Verdinelli}(1995)}]{chaloner_bayesian_1995}%
  \BibitemOpen
  \bibfield  {author} {\bibinfo {author} {\bibfnamefont {K.}~\bibnamefont
  {Chaloner}}\ and\ \bibinfo {author} {\bibfnamefont {I.}~\bibnamefont
  {Verdinelli}},\ }\href@noop {} {\bibfield  {journal} {\bibinfo  {journal}
  {Statistical Science}\ }\textbf {\bibinfo {volume} {10}},\ \bibinfo {pages}
  {273} (\bibinfo {year} {1995})}\BibitemShut {NoStop}%
\bibitem [{\citenamefont {Granade}\ \emph {et~al.}(2012)\citenamefont
  {Granade}, \citenamefont {Ferrie}, \citenamefont {Wiebe},\ and\ \citenamefont
  {Cory}}]{granade_robust_2012}%
  \BibitemOpen
  \bibfield  {author} {\bibinfo {author} {\bibfnamefont {C.~E.}\ \bibnamefont
  {Granade}}, \bibinfo {author} {\bibfnamefont {C.}~\bibnamefont {Ferrie}},
  \bibinfo {author} {\bibfnamefont {N.}~\bibnamefont {Wiebe}}, \ and\ \bibinfo
  {author} {\bibfnamefont {D.~G.}\ \bibnamefont {Cory}},\ }\href {\doibase
  10.1088/1367-2630/14/10/103013} {\bibfield  {journal} {\bibinfo  {journal}
  {New Journal of Physics}\ }\textbf {\bibinfo {volume} {14}},\ \bibinfo
  {pages} {103013} (\bibinfo {year} {2012})}\BibitemShut {NoStop}%
\bibitem [{\citenamefont {Granade}\ \emph {et~al.}(2017)\citenamefont
  {Granade}, \citenamefont {Ferrie}, \citenamefont {Hincks}, \citenamefont
  {Casagrande}, \citenamefont {Alexander}, \citenamefont {Gross}, \citenamefont
  {Kononenko},\ and\ \citenamefont {Sanders}}]{granade_qinfer_2017}%
  \BibitemOpen
  \bibfield  {author} {\bibinfo {author} {\bibfnamefont {C.}~\bibnamefont
  {Granade}}, \bibinfo {author} {\bibfnamefont {C.}~\bibnamefont {Ferrie}},
  \bibinfo {author} {\bibfnamefont {I.}~\bibnamefont {Hincks}}, \bibinfo
  {author} {\bibfnamefont {S.}~\bibnamefont {Casagrande}}, \bibinfo {author}
  {\bibfnamefont {T.}~\bibnamefont {Alexander}}, \bibinfo {author}
  {\bibfnamefont {J.}~\bibnamefont {Gross}}, \bibinfo {author} {\bibfnamefont
  {M.}~\bibnamefont {Kononenko}}, \ and\ \bibinfo {author} {\bibfnamefont
  {Y.}~\bibnamefont {Sanders}},\ }\href {\doibase 10.22331/q-2017-04-25-5}
  {\bibfield  {journal} {\bibinfo  {journal} {Quantum}\ }\textbf {\bibinfo
  {volume} {1}},\ \bibinfo {pages} {5} (\bibinfo {year} {2017})},\ \bibinfo
  {note} {arXiv: 1610.00336}\BibitemShut {NoStop}%
\bibitem [{\citenamefont {Huan}\ and\ \citenamefont
  {Marzouk}(2013)}]{huan_simulation-based_2013}%
  \BibitemOpen
  \bibfield  {author} {\bibinfo {author} {\bibfnamefont {X.}~\bibnamefont
  {Huan}}\ and\ \bibinfo {author} {\bibfnamefont {Y.~M.}\ \bibnamefont
  {Marzouk}},\ }\href {\doibase 10.1016/j.jcp.2012.08.013} {\bibfield
  {journal} {\bibinfo  {journal} {Journal of Computational Physics}\ }\textbf
  {\bibinfo {volume} {232}},\ \bibinfo {pages} {288} (\bibinfo {year}
  {2013})}\BibitemShut {NoStop}%
\bibitem [{\citenamefont {Dushenko}, \citenamefont {Ambal},\ and\ \citenamefont
  {McMichael}(2020)}]{dushenko_sequential_2020}%
  \BibitemOpen
  \bibfield  {author} {\bibinfo {author} {\bibfnamefont {S.}~\bibnamefont
  {Dushenko}}, \bibinfo {author} {\bibfnamefont {K.}~\bibnamefont {Ambal}}, \
  and\ \bibinfo {author} {\bibfnamefont {R.~D.}\ \bibnamefont {McMichael}},\
  }\href {\doibase 10.1103/PhysRevApplied.14.054036} {\bibfield  {journal}
  {\bibinfo  {journal} {Physical Review Applied}\ }\textbf {\bibinfo {volume}
  {14}},\ \bibinfo {pages} {054036} (\bibinfo {year} {2020})}\BibitemShut
  {NoStop}%
\bibitem [{\citenamefont {McMichael}, \citenamefont {Blakley},\ and\
  \citenamefont {Dushenko}(2021)}]{mcmichael_optbayesexpt_2021}%
  \BibitemOpen
  \bibfield  {author} {\bibinfo {author} {\bibfnamefont {R.~D.}\ \bibnamefont
  {McMichael}}, \bibinfo {author} {\bibfnamefont {S.~M.}\ \bibnamefont
  {Blakley}}, \ and\ \bibinfo {author} {\bibfnamefont {S.}~\bibnamefont
  {Dushenko}},\ }\href {\doibase 10.6028/jres.126.002} {\bibfield  {journal}
  {\bibinfo  {journal} {Journal of Research of the National Institute of
  Standards and Technology}\ }\textbf {\bibinfo {volume} {126}},\ \bibinfo
  {pages} {126002} (\bibinfo {year} {2021})}\BibitemShut {NoStop}%
\bibitem [{\citenamefont {Shor}(1994)}]{shor_algorithms_1994}%
  \BibitemOpen
  \bibfield  {author} {\bibinfo {author} {\bibfnamefont {P.~W.}\ \bibnamefont
  {Shor}},\ }in\ \href {\doibase 10.1109/SFCS.1994.365700} {\emph {\bibinfo
  {booktitle} {Proceedings 35th {Annual} {Symposium} on {Foundations} of
  {Computer} {Science}}}}\ (\bibinfo {year} {1994})\ pp.\ \bibinfo {pages}
  {124--134}\BibitemShut {NoStop}%
\bibitem [{\citenamefont {Kitaev}(1996)}]{kitaev_quantum_1996}%
  \BibitemOpen
  \bibfield  {author} {\bibinfo {author} {\bibfnamefont {A.~Y.}\ \bibnamefont
  {Kitaev}},\ }in\ \href {http://www.eccc.uni-trier.de/report/1996/003/} {\emph
  {\bibinfo {booktitle} {Electr. {Coll}. {Comput}. {Complex}.}}}\ (\bibinfo
  {year} {1996})\ pp.\ \bibinfo {pages} {TR96--003}\BibitemShut {NoStop}%
\bibitem [{\citenamefont {Hayes}\ and\ \citenamefont
  {Berry}(2014)}]{hayes_swarm_2014}%
  \BibitemOpen
  \bibfield  {author} {\bibinfo {author} {\bibfnamefont {A.~J.~F.}\
  \bibnamefont {Hayes}}\ and\ \bibinfo {author} {\bibfnamefont {D.~W.}\
  \bibnamefont {Berry}},\ }\href@noop {} {\bibfield  {journal} {\bibinfo
  {journal} {Physical Review A}\ }\textbf {\bibinfo {volume} {89}},\ \bibinfo
  {pages} {013838} (\bibinfo {year} {2014})}\BibitemShut {NoStop}%
\end{thebibliography}%

\newpage
\begin{center}
    FIGURE 1
\end{center}
\vspace{1in}
\begin{figure}[h]
    \centering
    \includegraphics[width=6.5in]{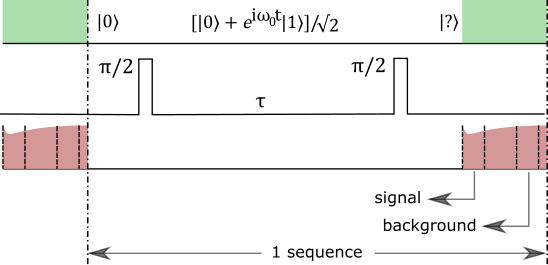}
\end{figure}

\newpage
\begin{center}
    FIGURE 2
\end{center}
\vspace{1in}
\begin{figure}[h]
    \centering
    \includegraphics[width=6.5in]{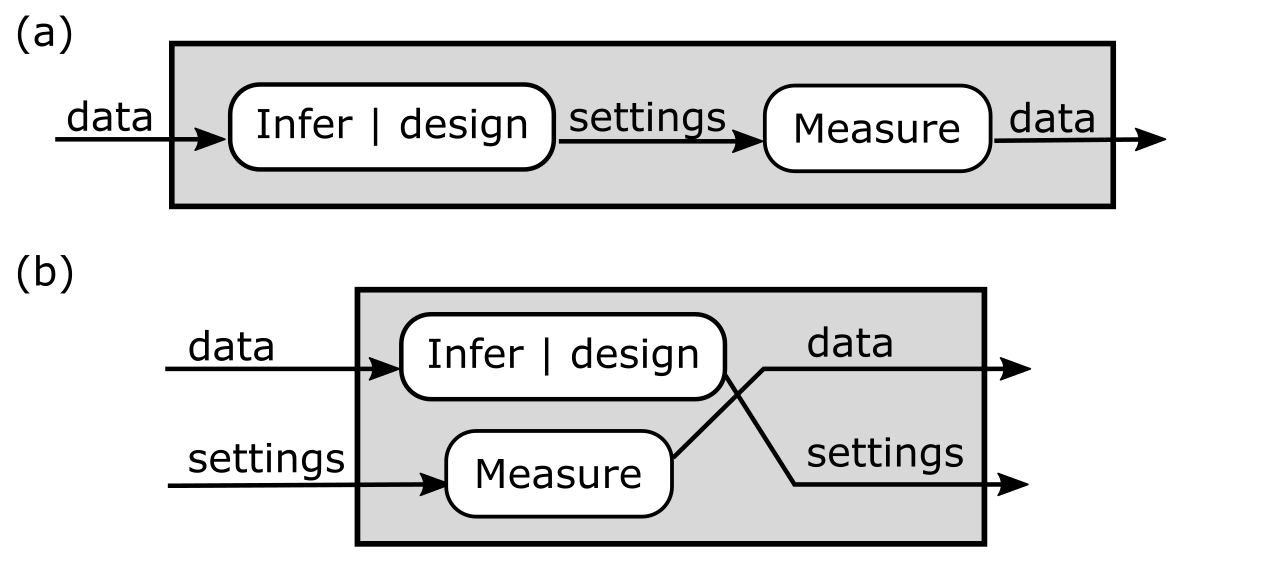}
\end{figure}

\newpage
\begin{center}
    FIGURE 3
\end{center}
\vspace{1in}
\begin{figure}[h]
    \centering
    \includegraphics[width=6.5in]{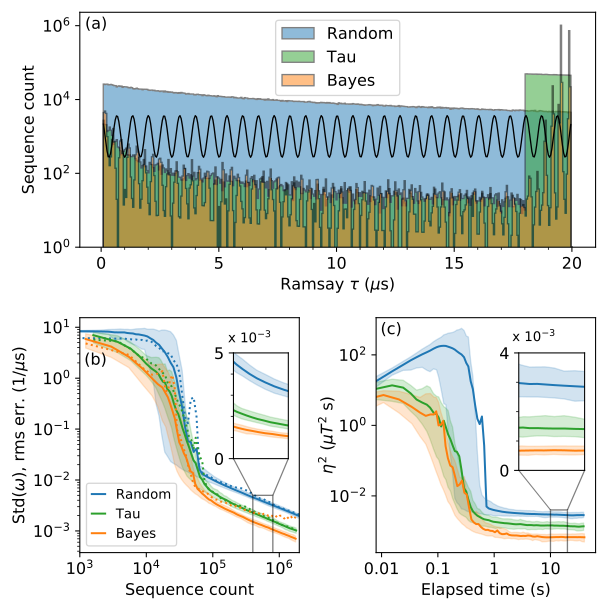}
\end{figure}

\newpage
\begin{center}
    FIGURE 4
\end{center}
\vspace{1in}
\begin{figure}[h]
    \centering
    \includegraphics[width=6.5in]{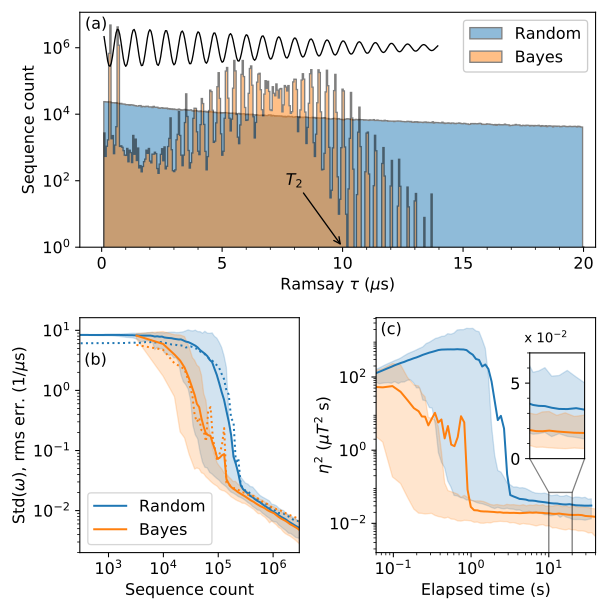}
\end{figure}

\newpage
\begin{center}
    FIGURE 5
\end{center}
\vspace{1in}
\begin{figure}[h]
    \centering
    \includegraphics[width=6.5in]{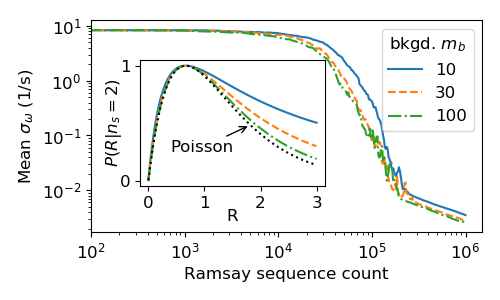}
\end{figure}

\end{document}